# Deep-Learning-Enabled Fast Optical Identification and Characterization of Two-Dimensional Materials

Bingnan Han[1,2,†], Yuxuan Lin[2,†,*], Yafang Yang[3], Nannan Mao[2,4], Wenyue Li[1], Haozhe Wang[2], Kenji Yasuda[3], Xirui Wang[3], Valla Fatemi[3], Lin Zhou[2], Joel I-Jan Wang[5], Qiong Ma[3], Yuan Cao[3], Daniel Rodan-Legrain[3], Ya-Qing Bie[3], Efrén Navarro-Moratalla[6], Dahlia Klein[3], David MacNeill[3], Sanfeng Wu[3], Hikari Kitadai[4], Xi Ling[4,7], Pablo Jarillo-Herrero[3,*], Jing Kong[2,5,*], Jihao Yin[1,*], Tomás Palacios[2,*]

[1] Image Processing Center, School of Astronautics, Beihang University, Beijing 100191, China

[2] Department of Electrical Engineering and Computer Science, Massachusetts Institute of Technology, Cambridge, MA 02139, USA

[3] Department of Physics, Massachusetts Institute of Technology, Cambridge, MA 02139, USA

[4] Department of Chemistry, Boston University, Boston, MA 02215, USA

[5] Research Laboratory of Electronics, Massachusetts Institute of Technology, Cambridge, MA 02139, USA

[6] Instituto de Ciencia Molecular, Universidad de Valencia, c/Catedrático José Beltrán 2, 46980 Paterna, Spain

[7] Division of Materials Science and Engineering, Boston University, Boston, MA02215, USA

[†] These authors contributed equally to this work.

[*]Correspondence to: (J. Y.) jihaoyin@buaa.edu.cn; (Y.L.) liny@mit.edu; (P. J. H.) pjarillo@mit.edu; (J. K.) jingkong@mit.edu; (T. P.) tpalacios@mit.edu

**Abstract**

Advanced microscopy and/or spectroscopy tools play indispensable role in nanoscience and nanotechnology research, as it provides rich information about the growth mechanism, chemical compositions, crystallography, and other important physical and chemical properties. However, the interpretation of imaging data heavily relies on the "intuition" of experienced researchers. As a result, many of the deep graphical features obtained through these tools are often unused because of difficulties in processing the data and finding the correlations. Such challenges can be well addressed by deep learning [1, 2, 3, 4]. In this work, we use the optical characterization of two-dimensional (2D) materials as a case study, and demonstrate a neural-network-based algorithm for the material and thickness identification of exfoliated 2D materials with high prediction accuracy and real-time processing capability. Further analysis shows that the trained network can extract deep graphical features such as contrast, color, edges, shapes, segment sizes and their distributions, based on which we develop an ensemble approach topredict the most relevant physical properties of 2D materials. Finally, a transfer learning technique is applied to adapt the pretrained network to other applications such as identifying layer numbers of a new 2D material, or materials produced by a different synthetic approach. Our artificial-intelligence-based material characterization approach is a powerful tool that would speed up the preparation, initial characterization of 2D materials and other nanomaterials and potentially accelerate new material discoveries.

**Introduction**

Research on two-dimensional (2D) materials has grown exponentially over the past decade[5, 6]. Hundreds of 2D materials have been isolated and studied [6, 7, 8, 9], offering a wide range of optical and electronics properties, including metals, semiconductors, insulators, magnets, and superconductors. The most widely used approach to obtain high-quality 2D crystals today is mechanical exfoliation [10, 11], followed by 2D crystal "hunting" under an optical microscope (Figure 1, left). This task is time-consuming and difficult especially for inexperienced researchers, but tedious for experienced researchers. There has, therefore, been growing interest in automating such a process[12]. An automatic optical identification and/or characterization system requires an algorithm that performs reliably for different materials with high accuracy, is fast enough for real-time processing, and is easily adaptable to different optical setups and different user requirements with minimal additional human efforts. However, existing identification approaches[13, 14, 15, 16, 17, 18] only exploited the information about the optical contrast (Figure S1) of the 2D crystals. As a result, they are often limited to specific types of 2D crystals and microscope conditions, which do not meet the aforementioned requirements.

In reality, optical microscopic (OM) images contain rich, often unused information other than optical contrast. These deep graphical features can be extracted through deep learning, especially semantic segmentation methods based on convolutional neural networks (CNNs) [1, 2, 3, 4]. In this work, an encoder-decoder semantic segmentation network[4] is configured for pixel-wise identification of OM images of 2D materials. We call our method 2D material optical identification neural network (2DMOINet) We demonstrate that this architecture can identify, in real time, various 2D materials in OM images regardless of variations in optical setups. This would free up tremendous amount of time for researchers working on 2D materials. Additionally, we find that the algorithm finds correlations between the OM images and physical properties of the 2D materials and can thereby be used to anticipate the properties of new, as-yet uncharacterized 2D crystals.

**Results**

**"2DMOINet" for 2D material identification**

Figure 1 illustrates the flow chart of our 2D material optical identification neural network (2DMOINet) method. Figures 1 (a) and (c) show 13 different 2D materials used for training, their crystal structures, photos of their bulk (three dimensional) source crystals [19] and representative OM images of exfoliated 2D crystallites (or "flakes") on top of 285 nm or 90 nm $SiO_2$/Si substrates. These materials are graphene/graphite, hexagonal boron nitride (hBN), the 2H-phase semiconducting transition-metal dichalcogenides (TMDs) $2H-MoS_2$, $2H-WS_2$, $2H-WSe_2$ and $2H-MoTe_2$, the 2H-phase metallic TMDs $2H-TaS_2$ and $2H-NbSe_2$, the 1T-phase TMD $1T-HfSe_2$, black phosphorous (BP), the metal trihalides $CrI_3$ and $RuCl_3$, and the quasi-one-dimensional (quasi-1D) crystal $ZrTe_5$. A total number of 817 OM images containing exfoliated flakes of these 13 materials and 100 background-only images were collected and labeled for training and testing. To make the training data representative of the typical variability of OM images, the training and test data were sampled from a collection of OM images generated by at least 30 users from 8 research groups with 6 different optical microscopes over a span of 10 years. The data were labeled with the material identifications and thicknesses in a pixel-by-pixel fashion with the help of a semi-automatic image segmentation program (Figure S3 and Methods). After color normalization and image resizing (Figure S2 and Methods), a training dataset of 3825 RGB images and a test dataset of 2550 images, both with the size of 224 by 224 pixels, were generated from these 917 OM images. During training, a random rotation data augmentation method which produce random positioning and orientation of images were utilized. Detailed information is shown in Methods. A stochastic gradient descent with momentum (SGDM) method[20] was used to train the weights in the convolutional filters of the 2DMOINet. The 2DMOINet (the schematic is shown in Figure 1 (d)) consists of a series of down-sampling layers (encoder) and a corresponding set of up-sampling layers (decoder) followed by a pixel-wise classifier. We selected a well-known network structure called VGG16 [21] as the encoder network in the 2DMOINet, and the detailed information about the network structure, the data generation and augmentation, as well as the network training process can be found in Figures 1 and S2, Table S1 and Methods section.

We first show that the trained 2DMOINet can be used to segment the OM images among 13 different exfoliated 2D materials and find the material identity and thickness of each flake. The performance can be visualized in Figure 1, Table 1 and Figure S4. The color maps in Figure 1 (e) are the typical label maps predicted by the trained 2DMOINet with the corresponding OM images in Figure 1 (c) as the input to the 2DMOINet. Figure S4 shows additional results of the test OM images, the ground-truth label maps (labeled semi-automatically by humans), as well as the predicted label maps. In addition to material separability, we labeled four materials (graphene/graphite, $2H\text{-}MoS_2$, $2H\text{-}WS_2$, and $2H\text{-}TaS_2$) with different thicknesses (1L for monolayer, 2-6L for bilayer to 6-layer, and >6 for greater than 6 layers) to verify the thickness differentiation capability, which is a particularly important task. The trained 2DMOINet is able to outline individual flakes from the background and distinguish both the material identities and thicknesses of the thirteen 2D materials with high success rate.

**Automatic 2D material searching system**

After a systematic optimization in terms of the network structure, the data augmentation approach, and the addition of background-only images (see Method for detailed information), we eventually obtained a trained 2DMOINet with high prediction accuracy, fast processing speed, as well as good generality that can be used for a fully automatic 2D material searching system. In the following, we discuss these three important aspects of our deep-learning-based optical identification approach in more details.

**(1) Accuracy**

Figure 2 presents the pixel-level (Figure 2 (a-e)) and the flake-level (Figure 2 (f-j)) confusion matrices of the test dataset (see Methods for more details). The diagonal elements are the success rate of each class, and the off-diagonal elements are the rate of misclassified pixels or flakes in the test OM images. The classification accuracies of both material identities (Figure 2 (a) and (f)) and thicknesses (Figure 2 (b-e) and (g-j)) are mostly above 70% and the mean class accuracy is 79.78%. Additional performance metrics are summarized in Table 1. The overall accuracy counted by pixels reaches 96.89%; and the mean intersection over union (IoU, defined as the intersection of the ground truth and the predicted region of a specific label over the union of them) is 58.78% for the training dataset. As will be discussed later, the accuracy of our network shown here is very promising especially for applications of automating the initial scanning and screening of exfoliated 2D materials.

Note that the calculated performance metrics of the 2DMOINet are likely an underestimate: after careful inspection of the label maps predicted by the 2DMOINet, we discovered a number of OM images in which the ground-truth was initially mislabeled, but predicted correctly by the 2DMOINet (Figure S5), which contributes tremendously to the thickness identification errors. This scenario is considered as a classification mistake in the above metrics. On the other hand, it is observed that many of the mistakes made by the network are due to the similarities between different materials. For example, misclassification rates among 2H-$MoS_2$, 2H-$WS_2$, 2H-$WSe_2$ and

2H-MoTe$_2$ are as high as 12%, which is probably a consequence of their similar crystal structures and optical properties. Although this type of misclassification is inevitable and detrimental to the prediction accuracies for material identity, it contains the information about the similarities between different materials which could be harnessed to construct material property predictors as will be discussed later. The third type of common mistake is that metal markers, tape adhesive residue and text labels in the OM images were misidentified as a 2D material (Figure S5). These non-2D material features were labeled as "background" together with the blank substrate in the ground-truth, but they have special structures and high color contrast relative to the substrate, thereby confusing the network. In a future version of the network, this may be solved by either further increasing the training dataset or introducing dedicated labels for these non-2D-material features. The former potential solution has been partially confirmed by our results. We observed a drop of the misclassification rate for the non-2D-material features if we compare the training results with the random-rotation augmentation method and additional background-only images, and those with the basic data augmentation method (Figure S6). The forth common mistake is inaccuracy in the profiles of the flakes. This usually happens when the profiles are very complex, or if the flakes are highly fragmentary (Figure S5). These mistakes are mainly due to the down-sampling of the encoder layers in the 2DMOINet, which inevitably drops the high-frequency spatial features of the images.

**(2) Speed**

We believe that the proposed deep learning algorithm is well suited to real-time processing according to the metrics given in Table 1. With our computing environment (see Methods), the training process for the VGG16 2DMOINet requires 30 hours with a GPU, whereas the testing speed can be as high as 2.5 frames per second (fps) using a CPU, and 22 fps using a GPU for 224-

by-224-pixel test images. This means the 2DMOINet, once properly trained, can be easily adapted to standard desktop computers and integrated with optical microscopes with automatic scanning stages for fast or even real-time identification.

As a proof-of-concept demonstration, we have integrated our 2D material identification algorithm to an optical microscope with a motorized XYZ stage. The motorized stage is able to scan across one sample (1.5 cm by 1.5 cm) and capture over 1,000 high-resolution OM images (1,636 by 1,088 pixels) within 50 minutes. These images can then be pre-processed (color normalization, resize) and fed to the trained 2DMOINet for their corresponding label maps in less than 15 minutes with a GPU-equipped computer. Figures S7, S8, and supplementary video 1 and 2 show the thickness identification results for an exfoliated graphene sample and an exfoliated $MoS_2$ sample. The frames outlined in red in Figures S7 and S8 are the identified monolayer and fewlayer samples. 9.45% and 8.25% of the total images are selected for the graphene and $MoS_2$ examples, respectively. These selected images can be further screened by human researchers within 5 minutes. Re-location of the appropriate flakes can then be implemented automatically for following studies. Additional information about this automatic 2D-material-searching system can be found in Methods. Furthermore, through the transfer learning technique discussed later, it is possible to further improve the thickness identification accuracy for a single material. This thickness identification tool has high identification accuracy and high throughput, and has already been utilized in daily laboratory research.

### (3) Robustness and generality

We would like to emphasize that our deep-learning-based optical identification approach is generic for identification of different 2D materials, and has good robustness against variations such as brightness, contrast, white balance, and non-uniformity of the light field (Figure S2, left). For this, we intentionally included OM images taken under different optical microscopes and different users as mentioned previously. As a comparison, existing optical identification methods are completely based on optical contrast of the 2D crystals[13, 14, 15, 16, 17, 18], and as a result, they are often specific to types of 2D crystals, conditions and configurations of the microscopes being used, image qualities, *etc*. In addition, optical contrast based methods would fail for harder problems in which the classes to be differentiated are not separable in the color space, such as identifying the materials in unlabeled optical images (Figure S1). Our results suggest that after training, the same 2DMOINet is able to identify both the material identities and their thicknesses among 13 different 2D materials with high accuracy and fast speed from OM images regardless of different conditions of the microscopes, which is fundamentally different and more practical as compared to optical-contrast-based approaches. Another important feature we discovered from our experiment is that the trained 2DMOINet was even able to identify the OM images that are out of focus (Figure S9). This further proves the robustness of our algorithm.

**Predicting physical properties of "unknown" materials** In the following, we first discuss how our 2DMOINet was able to capture deep graphical features. Based on this, we then constructed an ensemble approach to provide predictions about physical properties of unknown 2D materials.

  **(1) Interpretation of the 2DMOINet**

To interpret how the 2DMOINet extracts features from 2D material OM images, we analyzed the output feature maps of all the layers in the trained network for the OM images in the test dataset as the inputs. As a demonstration, we used a typical image of graphite/graphene (shown in Figure S13 (a)) as the input. The corresponding convolutional feature maps of all the layers in the encoder, decoder, and output sections of the 2DMOINet are summarized in Figure S13 (b)-(d). Taking the convolutional feature maps in the encoder as an example (Figure S13 (b)), we can clearly see that

Depth=1 feature maps are highly correlated to color and contrast information. In this shallow layer, the background and monolayer region of graphene are not easily separable because of the weak contrast between the two classes, whereas multilayer graphene region is already quite distinguishable. In Depth=2 feature maps, more boundary characteristics are detected, and the edges of graphene monolayer regions start to stand out in some of the feature maps. With the increase of the depth, the size of each feature map becomes smaller because of the pooling layers in the network, and the receptive field of each convolutional kernel (the area in the input image each neuron in the current layer can respond to) becomes relatively larger, which leads to a higher level abstraction of the global graphical features. For instance, Figure S13 (e) displays the most prominent feature map (channel #186) of the Depth=5 encoder layer with the largest activation value. It is observed that this feature map is highly correlated to the monolayer graphene region. By feeding the network with more test images as summarized in Figure S22, we further confirmed that channel #186 is sensitive to pink/ purple flakes with large flake sizes, smooth edges and regular shapes.

After further analysis on the 512 channels of the Depth=5 encoder layer with more test images randomly chosen from our database, we concluded that the trained 2DMOINet is able to capture deep and subtle graphical features that were overlooked by previously reported optical contrast based approaches [13, 14, 15, 16, 17, 18]. Many of the deep graphical features reflect in part the physical properties of the 2D materials. To illustrate this, we select 21 easily interpreted channels and discuss their associated graphical features and the related physical properties as summarized in Figure 3, Figures S14-34, Table S5 and S6. We divide the graphical features captured by these channels into four broad categories: (1) contrast or color, (2) edge or gradient, (3) shape, and (4) flake "size". Figure 3 shows the heat maps of several channels that belong to each category. In particular, channels #186, #131 and #230 under the "contrast/color" category (Figure 3 (b)) are sensitive to flakes with purple/pink, yellow/green/grey and bright green/blue colors, respectively; channels #87 and #488 under the "edge" category (Figure 3 (c)) reveal smooth all-direction edges and bottom edges, respectively; channels #52, #436 and #279 under the "shape" category (Figure 3 (e)) are indicators of shapes with edges at acute angles, slender shapes, and one-dimensional (1D) straight wires; and channels #230 and #484 under the "flake size" category capture large and small/fragmentary flakes. Table S6 and Figures S14-34 provide more OM images, their corresponding heat maps as well as the extracted graphical features of the 21 channels. Note that some channels can only respond to images that meet a combinational criterion under multiple categories, whereas some channels can be sensitive to several scenarios. For example, channel #87 only shows high intensities in the heat map around the smooth edges of purple flakes (Figure S16), and channel #312 can be used to identify both non-uniform, thick flakes and uniform, thin, purple/pink 1D wires (Figure S27).

To confirm that the above analysis about the deep graphical feature maps are generic for our problem but not specific to any individual 2DMOINet, we repeated the training procedure for the 2DMOINet with different initialization, random permutation of the dataset, and different data augmentation strategies. 16 training runs were performed separately with the metrics and pixel-level confusion matrices shown in Table S4 and Figures S36-50. Similar behaviors are manifested in all these 2DMOINets. Moreover, we can easily identify similar functionalities among the Depth=5 channels in these models. In Figure S51, four pairs of channels that belong to two independently trained networks (network #1 and network #2 in Table S4) are displayed side by side. Although the channel indices are different in each pair, they all have similar feature activation behavior on the same input OM images. In particular, channel #153 in network #1 and channel

#227 in network #2 correspond to low-contrast purple/pink flakes with smooth edges; channel #13 in network #1 and channel #433 in network #2 correspond to left edges of low-contrast purple/pink flakes; channel #490 in network #1 and channel #433 in network #2 correspond to 1D wires; and channel #76 in networ #1 and channel #490 in network #2 correspond to fragmentary/small flakes.

Another interesting observation regarding the generality of the graphical feature found by our algorithm is that the 2DMOINet is able to learn additional features in the Depth=5 channels in order to compensate the artifacts induced by the random rotation data augmentation strategy. In Figure S52 (b) and (c), it is clearly observed that a circular region appears in the feature map using random rotation method, and no such pattern emerges in basic augmentation method. This difference is caused by a padding procedure used during image rotation in every epoch. However, we find that the network amends this "error" and predict a correct map as shown in Figure S52 (d). Furthermore, it is clearly seen that this artifact has been mostly removed in the output of the graphene channels in the final ReLU layer of the network using random rotation augmentation method as shown in Figure S52 (e). We also found that additional channels that are only sensitive to the perimeters of the squared images are generated by the models with random rotation augmentation (channels #156, #402 and #425 in Table S6, see examples in the inset of Figure S52 (b) and Figures S20, S29, S30). These features are much less prominent in the model with the basic data augmentation strategy. Therefore, we believe that these channels related to image perimeters are responsible for the compensation of the circular artifacts induced by the random rotation augmentation.

**(2) Predicting material properties through an ensemble approach**

The above feature map analysis has provided a better understanding about how deep graphical features can be extracted by the 2DMOINet for more accurate and generic optical identification of exfoliated 2D materials. However, the algorithm is not limited to this particular task, and we found that it can be used for more advanced optical characterization tasks such as the prediction of material properties. For this, the graphical features captured by the network are correlated to the optical and mechanical properties of the material. As shown schematically in Figure 3 (a), the contrast/color and the edge features are determined by the optical response of the material, which reflect the electronic band structures and the thicknesses of the flakes. In addition, because the samples were made through mechanical exfoliation, the typical distributions of flake shapes and sizes heavily depends on the mechanical properties of the materials, such as the crystal symmetry, mechanical anisotropy and the exfoliation energy (Figure 3 (d)). We can thus use the network trained with the knowledge of the 13 materials and do further statistical analysis to predict properties of materials not present in the training set.

One simple approach to implement the material property predictor is to take the "similarity" vectors of each material from the confusion matrix (each row of the matrix) and decompose the vectors to a set of specifically designed base vectors. To improve the fidelity of this simple and relatively weak predictor, an ensemble approach is employed[22]. We performed such decomposition operation on the 16 2DMOINets (Table S4) that were trained independently, and took the average of these predictors as the final prediction. In this way, the projected values can give us a probabilistic prediction of the physical property of interest (more details are given in the Methods section).

As a proof-of-concept demonstration, we fed the trained network with OM images of 17 2D materials that were unknown to the network during the training stage. The extended confusion matrices containing the 13 trained materials and the 17 untrained materials is shown in Figure 36-50, and the physical properties of these 2D materials are summarized in Table S5. As we can see, different vector components in the similarity vectors (or columns of the extended confusion matrix) have distinct values for each of the untrained materials, from which we can immediately summarize some qualitative patterns. For example, GaS, $CrI_3$, $CrBr_3$ and $MnPS_3$ in the untrained material group shows high similarity to hBN in the trained material group, which matches the fact that these materials are wide-bandgap semiconductors or insulators with the band gaps higher than 2.5 eV and are mostly transparent in the infrared, red and green spectral ranges. As another example, 1T'-$MoTe_2$ and Td-$WTe_2$ in the untrained material group are predicted to be similar to 1T-$HfSe_2$ in the trained material group, which is in accordance with the similar crystal structure of these materials.

For a more quantitative analysis, we selected two different predictors that are associated with the band gaps and the crystal structures of the material, and the projected mean values and their standard deviations of each material based on these two predictors and the 16 2DMOINets are plotted into a histogram with an error bar as summarized in Figure 4 (a) and (b), respectively. Clear correlations between the projected values and the true physical parameters (band gap in Figure 4 (a), and crystal structure in Figure 4 (b)) are found. We can thus use the projected values as an indication of the probability of the physical property of interest of an unknown material belonging to each class (represented by each base vector of the corresponding predictor). Note that there are also misclassified instances, which we believe can be improved by either constructing a more complex material property predictor, or expanding the training data set in terms of both the number of images and the number of materials. This method can also be used for systematic studies of other factors such as the effect of different mechanical exfoliation techniques, bulk crystal qualities, and so on.

**Reconfiguring "2DMOINet" for other optical characterization applications**

Finally, we show that the trained 2DMOINet can be adapted for different applications through transfer learning. The basic idea is to use the trained 2DMOINet as the initialization for the new training problem rather than a random initialization. With this approach, we are able to train the 2DMOINet for new optical identification/characterization problems with a good balance between prediction accuracy and computational/data cost. Here we use OM images of graphene synthesized by chemical vapor deposition (CVD) and exfoliated Td-WTe$_2$ as two examples. Figures 5 (a), S52 and 5 (c), S53 display the test images, the ground truth label maps as well as the corresponding prediction results after the trainings on the CVD graphene and the exfoliated Td-WTe$_2$ datasets, respectively, with the transfer learning approach. As we can see, the prediction results match the ground truths very well. To compare the pre-training transfer learning approach with the conventional approach with random initialization, we plot the global test accuracy of networks trained with both approaches as a function of the number of OM images in the CVD-graphene dataset as shown in Figure 5 (b). Representative test images and their predictions can be found in Figure S53. With the pre-training approach, we were able to achieve 67% global accuracy with only 60 training images, whereas at least 240 images are required for the conventional random initialization approach to reach comparable accuracy. In addition, we also showed in Figure 5 (d) the pixel-level confusion matrix of the transfer-learning 2DMOINet on the Td-WTe$_2$ dataset. Because this newly trained 2DMOINet is dedicated to the thickness prediction for only one material, it is able to differentiate different thicknesses with much better accuracy (91% mean class accuracy) and finer thickness divisions.

**Discussion**

In summary, we develop a deep learning algorithm for the optical characterization of 2D materials and the extraction of deep graphical features from optical microscopic images that can be used for anticipating material properties. After training, the neural network can achieve rapid characterization of material identities and their thicknesses with high accuracy. A fully automated system utilizing this algorithm is developed, which is able to free up tremendous amount of time for researchers. In addition, a systematic analysis was made to understand how the network captures deep graphical features such as color, contrast, edges, shapes and flakes sizes from the optical images. Based on this, an assemble approach derived from the trained neural network can be used for the prediction of physical properties of the materials. We also demonstrate that the trained network can be adapted for different optical characterization applications with minimal additional training through a transfer learning technique. The proposed methodology can potentially be extended for identification and understanding other morphological or spectroscopic data of diverse nanomaterials.

**Methods**

**Constructing, Training and Testing the 2DMOINet**

(1) Network structure

The 2DMOINet-based optical identification process is summarized in Figure 1 and Figure S2. The 2DMOINet consists of a series of down-sampling layers (encoder) and a corresponding set of up-sampling layers (decoder) followed by a pixel-wise classifier. As an end-to-end network, the 2DMOINet can predict labels of 2D materials at the pixel level, and the size of the output label map is exactly the same as the input optical microscope image. This can help us directly identify the material identities and the thicknesses of individual 2D material flakes. A well-known network structure VGG16 [21] is selected as the basis of the encoder network in the 2DMOINet (except for the depth-dependence study). . Table S1 summarizes the 2DMOINet structure based on the

VGG16 network. The last three fully-connected layers are dropped. The encoder contains a stack of convolutional layers which have 3 by 3 receptive field and pixel stride 1, followed by a batch normalization and a nonlinear activation (rectified linear unit, ReLU) layer. Then a max-pooling layer with a 2 by 2 window and stride 2 is applied for the image down-sampling. The decoder net and the encoder net are symmetric. The only difference between them is that in decoder we use an up-sampling layer to replace max-pooling layer. The indices in up-sampling layers are grabbed directly from the indices of the corresponding max-pooling layers. In this way, the locations of the pooling layers are memorized and recovered in the up-sampling layers, which improves the spatial accuracy of the network. Finally, a soft-max classifier is added at the end of the network for pixel-wise classifications. The output label maps have the same dimension as the input OM images. As for the depth-dependence study, we fixed the number of convolutional kernels in each convolutional layer and the number convolutional layers in each depth to be 64 and 2, respectively.

(2) Data generation

There are 3 steps to generate the pixel-wise labeled dataset for the training of the 2DMOINet: labeling, color normalization, and data augmentation.

To generate pixel-wise labeled OM images for the training and testing, we used a semi-automatic graph-cut method implemented by MATLAB. Graph-cut is a traditional semantic segmentation method based on graph theory[23]. Although the initial segmentation performance of this method is poor, we can promote the performance by adding human assistance. By loosely drawing the foreground regions and the background regions, the algorithm can find the boundary of the segment of interest with good accuracy. Figure S3 demonstrate the labeling procedure under human-assisted graph-cut method.

We select thirteen 2D materials as experiment samples. They are graphene/graphite, hexagonal boron nitride (hBN), 2H-MoS$_2$, 2H-WS$_2$, 2H-WSe$_2$, 2H-MoTe$_2$, 2H-TaS$_2$, 2H-NbSe$_2$, 1T-HfSe$_2$, black phosphorous (BP), CrI$_3$, RuCl$_3$ and ZrTe$_5$. About 100 images for each of Graphene/Graphite, 2H-MoS$_2$, 2H-WS$_2$ and 2H-TaS$_2$ were labeled with three classes ("monolayer", "fewlayer (2-6L)" and "multilayer (>6L)") to demonstrate the thickness identification capabilities, and about 50 images for each of the other 9 materials were labeled with only single classes. Moreover, to improve the robustness of our network, 100 background-only images are chosen which contain unbalanced light conditions or tape residuals. Therefore, the total number of classes (including background) is 22, and the total number of labeled OM images is 917. To partially reduce the color-related variations of the OM images because of different setups and user preferences, a color normalization was performed on all the images. We first converted the RGB images into the Lab color space, and the following transformation was applied to all the pixels: $L \leftarrow 30L/L_{ref}$, $a \leftarrow a-a_{ref}$, and $b \leftarrow b-b_{ref}$, where $L_{ref}$, $a_{ref}$ and $b_{ref}$ are the Lab values of the background obtained by finding the median $L$, $a$ and $b$ of each image. The resulting Lab images were then converted back to RGB images.

After labeling and color normalization, we selected 90% of the original labeled images as the training dataset, and 10% as the test dataset. The original images were first resized and chopped into 224 by 224 pixel images (450 for multi-classes materials and 225 for single-class materials for images selected for training; 300 for multi-classes materials and 150 for single-class materials for images selected for testing), so we obtain 3825 images (4275 if background images are added) in the basic training dataset, and 2,550 images in the test dataset.

In this paper, we select three different strategies for data augmentation, that is, basic data augmentation, color augmentation and random rotation augmentation. In the basic data augmentation, each chopped image was flipped (horizontal and vertical) and rotated (by 0°, 90°, 180° and 270°) to generate 6 augmented images, so the size of the training dataset is 22950. In color augmentation, we choose brightness ($I = (R+G+B)/3$), contrast ($C = [R\text{-}128, G\text{-}128, B\text{-}128]$) and color balance ($CB_R=R\text{-}128$; $CB_G=G\text{-}128$; $CB_B=B\text{-}128$) as transformation variables. 12 additional images are generated by multiplying these 5 variables with 5 different factors randomly drawn from a uniform distribution in the span of [0.9, 1.11]. The size of the training dataset for color augmentation is 49725. The third data augmentation method is random rotation. In each epoch, the rotation angle and mirror operations of images are set randomly. The size of the training dataset is still 3825 (without background-only images) or 4275 (with background-only images).

(3)    Training and testing

The training and testing of the 2DMOINet were implemented in *MATLAB R2018b* with the help of the *Deep Learning Toolbox*, the *Parallel Computing Toolbox*, the *Computer Vision Toolbox* and the *Image Processing Toolbox*. The training and testing were performed using a desktop computer equipped with a CPU (Intel(R) Core (TM) i7-8700K @ 3.70GHz, 32.0GB RAM) and a GPU (NVIDIA GeForce GTX 1080 Ti, 11 GB GDDR5X). The stochastic gradient descent with momentum (SGDM) method[20] was used to find the weights in the convolutional filters of the 2DMOINet during the training process. To compensate for the imbalanced numbers of pixels for different classes (for example, the "background" labels take more than 85% of areas in most images), class weightings based on inverse frequencies were used in the soft-max classifier. We set 20 epochs for the training process when using the basic and color data augmentation method, with the learning rate of $10^{-2}$ initially and $10^{-3}$ after 10 epochs. When using the rotation data augmentation method, we set 250 epochs with the learning rate of $10^{-2}$ initially, and $10^{-3}$ after 150 epochs. Using above parameters, we first test the performance of different data augmentation methods in Table S3, then we present the performance of 14 different training iterations with the random rotation augmentation method in Table S4. Networks #3-9 are trained using 4275 images (with additional background-only images), while networks #10-16 are trained using 3825 images. The overall performance of these 14 networks are summarized in Table S6. Network #5 is used in our final algorithm if not stated otherwise.

**Optimization of the 2DMOINet**

To optimize the performance of the 2DMOINet, we first investigated the network with different depths (number of pooling layers, as indicated in Figure 1) and tested their performance on the same data set. In this experiment, we fixed the number of convolutional kernels in each convolutional layer and the number of convolutional layers in each depth to be 64 and 2, respectively, rather than varying them gradually as the layer becomes deeper, as in the case of the VGG16 structure. Table S2 shows the metrics for the networks with different depths. As we can see, Depth=5 network results in the best global accuracy, mean class accuracy and mean IoU among all networks. In Figure S10, we present the training loss and the training accuracy under different network depths during the training process. The networks with depths from 1 to 6 can all reach 80% pixel-by-pixel global accuracy quickly within the first several epochs. This corresponds to the successful differentiation between the background and the foreground (2D material flakes). After that, the Depth=4 and Depth=5 network continue to reduce the loss and improve the training accuracy, while other networks only show limited training progress in terms of differentiating the material identities and thicknesses. Figure S11 and S12 are examples of the label maps predicted by networks with different depths. We can clearly see that the Depth=5 network results in the best prediction accuracy, while the other networks could outline the flakes from the background, but fail to differentiate among the classes (thicknesses and material identities). This justifies the use of the modified VGG16 structure as the encoder of the 2DMOINet given that the VGG16 structure is similar to the Depth=5 network structure in the depth-dependent study here.

Second, we designed control experiments to evaluate three different data augmentation strategies: (1) basic (translation, mirror, 4-fold rotation); (2) color (brightness, contrast, color balance); and (3) random rotation (arbitrary angle rotation applied on each epoch during the training). Details about these data augmentation strategies can be found in Methods. Five trainings were performed separately on the same VGG16 2DMOINet architecture but with different data augmentation strategies. The training results as displayed in Table S3 suggest that (1) all the data augmentation strategies are able to improve the prediction accuracy, mainly because the training dataset is expanded; (2) the color augmentation only provides limited improvement, but with a penalty of much longer training time because of the 13 times expansion of the training dataset; (3) the mean class accuracies, especially for different thicknesses are decreased with color augmentation, which may be attributed to the loss of color/contrast information; and (4) the random rotation augmentation gives rise to the best performance because much more variations can be "seen" by the 2DMOINet within a manageable timeframe of training. Therefore, we only implemented the random rotation augmentation strategy in our final optical identification algorithm.

Third, we also included additional OM images without any 2D flakes (background only) into the training dataset and evaluated the training accuracy. For this, we performed trainings with and without the background-only images, and the results are summarized in Table S4. Although the addition of extra background-only images gives rise limited improvements of the overall accuracies, it does help to reduce the misclassification rates for the high-contrast non-2D-material features such as tape residuals, metal marks, *etc*. as shown in Figure S6.

Based on the above analysis, we finally chose the VGG16-based 2DMOINet, the random rotation augmentation strategy, and the addition of background-only OM images as the optimized model (network #5 in Table S4).

**Confusion Matrices and Material Property Predictors**

The confusion matrices (Figure 2, Figure 5 (d) and Figure S36-S50) were obtained by using the OM images in the test dataset as the input of the trained 2DMOINet and comparing the corresponding output label maps with the ground truth label maps. For example, the element on $i$-th row and $j$-th column corresponds to the fraction of the $i$-th class that are labeled as the $j$-th class by the 2DMOINet. We also present two types of confusion matrices: the pixel-level confusion matrices (Figures 2 (a)-(e), 5 (d), and S35-S49) and the flake-level confusion matrices (Figure 2 (f)-(j)). For the pixel-level confusion matrices, the matrix elements are the ratio counted pixel by pixel, whereas the flake-level confusion matrices take the majority label of all the pixels in each segmentation, or "flake", as the label of the flake and calculated the fraction based on the flake labels. Note that the flake labels ignored any segmentations with fewer than 50 pixels, because they are either fractures on the edges of the actual 2D crystal flakes, or some non-uniform regions on the background.

To demonstrate the 2DMOINet's capability of predicting physical properties of unknown 2D materials, we also fed the trained 2DMOINet with additional OM images of new 2D materials that were not used in the training dataset, and calculated the "extended" confusion matrix as shown in Figure S36-S50. The new materials include 2H-$MoSe_2$, 1T'-$MoTe_2$, Td-$WTe_2$, $ReS_2$, $SnS_2$, $SnSe_2$, GeSe, SnSe, GaS, $CrCl_3$, $CrBr_3$, $MnPS_3$, $FePS_3$, $TiS_3$, $ZrS_3$, $Bi_4I_4$ and $Ta_2Se_8I$. The row vectors in the extended confusion matrix can be used to characterize how similar the physical properties of one material are to the 13 known materials. Therefore, we can use these "similarity" vectors to extract information about physical properties of a new material. A very simple way of constructing such physical property predictors is to project the "similarity" vectors in a set of base vectors that are correlated to the physical property of interest. We can define each base vector as the average of the vectors of the known materials that have the same value or range of the physical property of interest, expressed as $\mathbf{v}_k = (\sum_{x \in \mathbf{M}_k} \mathbf{v}_x)/\|\mathbf{M}_k\|$, where $\mathbf{v}_k$ is the base vector of the $k$-th class in the material property predictor; $\mathbf{v}_x$ is the base vector of the material $x$ in the training set; $\mathbf{M}_k$ is the subset of materials in the training set that have matched criteria of the physical properties in the $k$-th class in the predictor; and $\|\mathbf{M}_k\|$ is the number of materials in the $\mathbf{M}_k$ subset. For the band gap predictor, the base vector subsets are $\mathbf{M}_1$={Graphene/Graphite, 2H-TaS$_2$, 2H-NbSe$_2$, ZrTe$_5$, BP, 2H-MoTe$_2$, 1T-HfSe$_2$, RuCl$_3$, CrI$_3$}, $\mathbf{M}_2$={2H-MoS$_2$, 2H-WSe$_2$, 2H-WS$_2$}, $\mathbf{M}_3$={hBN}; for the crystal structure predictor, the base vector subsets are $\mathbf{M}_1$={Graphene/Graphite, hBN, 2H-MoS$_2$, 2H-WSe$_2$, 2H-WS$_2$, 2H-MoTe$_2$, 2H-TaS$_2$, 2H-NbSe$_2$}, $\mathbf{M}_2$={1T-HfSe$_2$}, $\mathbf{M}_3$={BP}, $\mathbf{M}_4$={RuCl$_3$, CrI$_3$}, $\mathbf{M}_5$={ZrTe$_5$}. An ensemble approach is used to improve the fidelity of these relatively simple predictors. We calculate the mean and the standard deviations of the outputs of these predictors for the 16 separately trained 2DMOINets (Table S4 and Figures S36-50) and use the mean values as the final outputs of the ensemble material property predictors. A summary of physical properties of the 2D materials being used in this study can be found in Table S5.

**Transfer Learning for Small Amount of Training Data**

In the first transfer learning experiment, we only labeled 5 OM images of CVD grown graphene. 4 OM images were used to generate the training dataset, and 1 OM image were used to generate the test dataset. 360 and 90 images (224 by 224 pixels) were generated in the training and test dataset with the basic data augmentation. We divided the graphene region into five classes based on its layer number (from monolayer to 5-layers). We varied the size of the training dataset from 30 to 360, and sampled randomly from the 360 images. In the second transfer learning experiment, a total number of 297 OM images of exfoliated Td-WTe$_2$ samples (including 48 background-only images) were used. Similar data augmentation and resizing strategy were applied before the training.

For the training of the transfer learning, the initial weights in the network before the training are taken from the pretrained 2DMOINet (trained with the 13 exfoliated 2D materials, network #5 in Table S4.) as compared to the conventional random initialization strategy. 200 epochs of training was made on both the pretrained 2DMOINet and the randomly initialized 2DMOINet. The learning rates are $10^{-2}$ and $10^{-3}$, respectively, for the first 100 epochs and the last 100 epochs.

**Automatic 2D material searching system**

We integrated our 2D material identification algorithm with an optical microscope (Nikon) with a motorized XYZ stage. We used an exfoliated graphene sample and an exfoliated MoS$_2$ sample with the size of 1.5 cm by 1.5 cm to test the system. it took nearly 50 minutes to finish the automatic scanning and produce over 1000 1636-by-1088-pixel images (graphene: 1862 images, MoS$_2$: 1406 images) per sample. Then color normalization and image chopping were applied to these images. To cover every part in one image, we first resized these images by 0.5, and then chopped into 12 squared images (224 by 224 pixels), which led to 22344 and 16872 images for the graphene and MoS$_2$ samples, respectively. Network #5 in Table S4 was used to output the predicted material label maps. It took less than 15 minutes to process all the images for each sample. The automatic scanning and the image processing/prediction can be executed either simultaneously or sequentially. Finally, we can set up the criteria (thickness, flake size, *etc.*) for the 2D materials of interest and re-locate the flakes for further inspection/characterization. In our examples, the criteria were set to be monolayer and fewlayer segments with >100 pixels. As a result, 9.45% and 8.25% of the total images are selected for the graphene and $MoS_2$ examples, respectively. These images can be inspected by human within 5 minutes. Supplementary video 1 and 2 summarized the automatic 2D material searching procedure for the graphene and the $MoS_2$ samples, respectively. Figure S7 and S8 are the corresponding overview images with the 2D flakes that meet our criteria outlined in red.

**Acknowledgement**

This material is based upon work sponsored in part by the U.S. Army Research Office through the Institute for Soldier Nanotechnologies, under cooperative agreement number W911NF-18-2-0048, AFOSR FATE MURI, grant no. FA9550-15-1-0514, National Natural Science Foundation of China (grant no. 41871240), the National Science Foundation grant 2DARE (EFRI-1542815), NSF DMR-1507806, and the STC Center for Integrated Quantum Materials, NSF Grant No. DMR-599 1231319. Work by the P. J-H. group was primarily funded by the DOE Office of Science, BES, under award DE-SC0019300 as well as the Gordon and Betty Moore Foundation via grant GBMF4541 B. H. gratefully acknowledges the financial support from China Scholarship Council.

**Author Contributions**

Y.L., Y.Y., B. H., J. Y., J. K. and T. P. conceived the experiment. B. H., Y. L., Y. Y. and W. L. performed the training and testing of the 2DMOINet. B. H., Y. Y., N. M. Y. L. and H. W. carried on the data labeling. B. H. and Y. L. did further analysis of the trained network. Y. Y., K. Y., and

# Figures and Captions

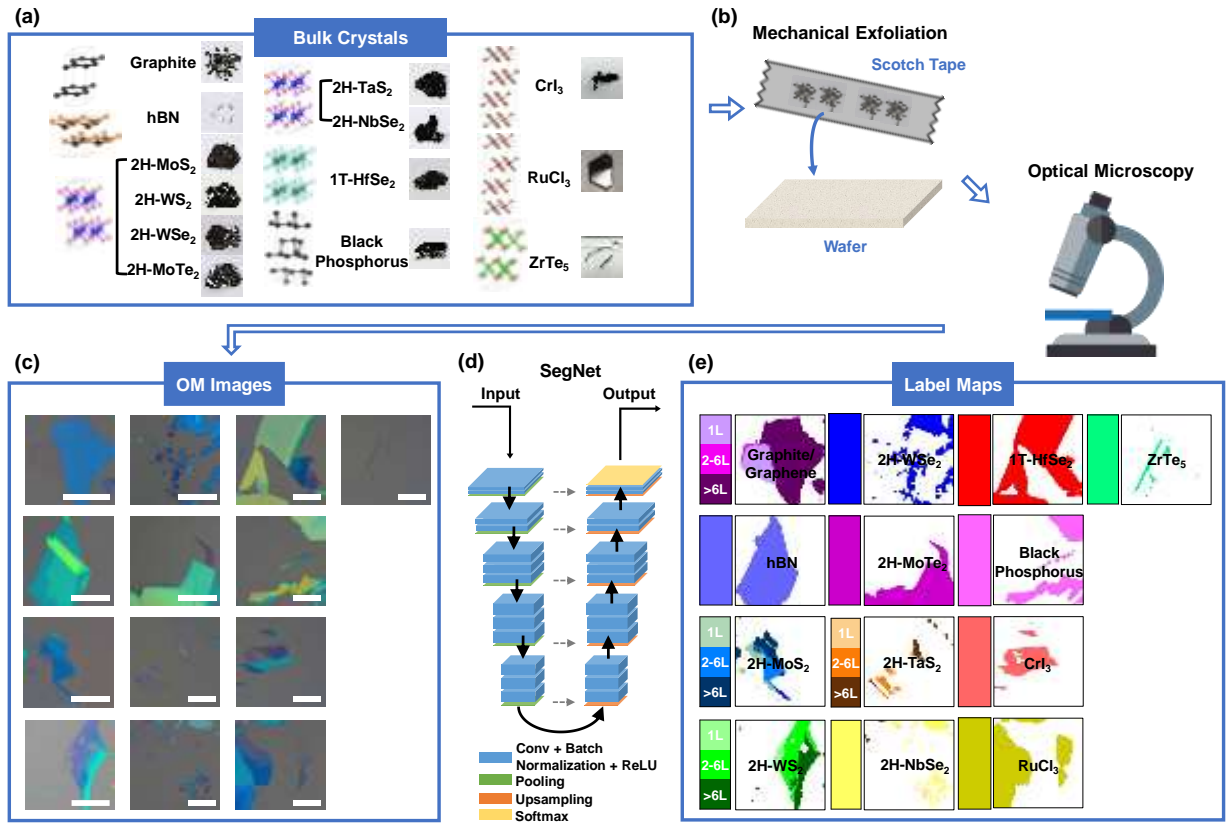

**Figure 1.** The flow chart of the proposed deep learning based optical identification method. We select 13 typical 2D materials (crystal structure and photographs [19] of the bulk crystals are shown in panel (a)). After mechanical exfoliation, the 2D flakes are randomly distributed on $SiO_2/Si$ substrates. We then use optical microscopes to take the images (b). Panel (c) shows representative optical microscopic (OM) images of the 13 materials. When inputting these images to the trained 2DMOINet (as shown in (d)), the label maps (e) will be predicted that segment individual 2D flakes and provide the labels (materials identities and thicknesses) of them. The 2DMOINet is composed of a series of convolutional layers, batch normalization layers, rectified linear unit (ReLU) layers (in blue), pooling (down-sampling) layers (in green), up-sampling layers (in orange), as well as a soft-max layer (in yellow) as the output layer. Note that the OM images displayed here

and in Figures 3 and 5 are the original high-resolution images. The images for the input of the 2DMOINet were down-sampled to 224 by 224 pixels, which can be found in Figures S4, S7-S20 and S22. Scale bars in (c), 20 μm.

**Figure 2.** Confusion matrices calculated from the test results. (a)-(e) are pixel-level confusion matrices, and (f)-(j) are flake-level confusion matrices (see Methods). (a) and (f) are for material identities. (b)-(e) and (g)-(j) are for thicknesses. In each confusion matrix, the diagonal terms are the success rate of the predicted classes, and the non-diagonal terms are the rate of misclassified pixels. The mean class prediction accuracies is 79.8%.

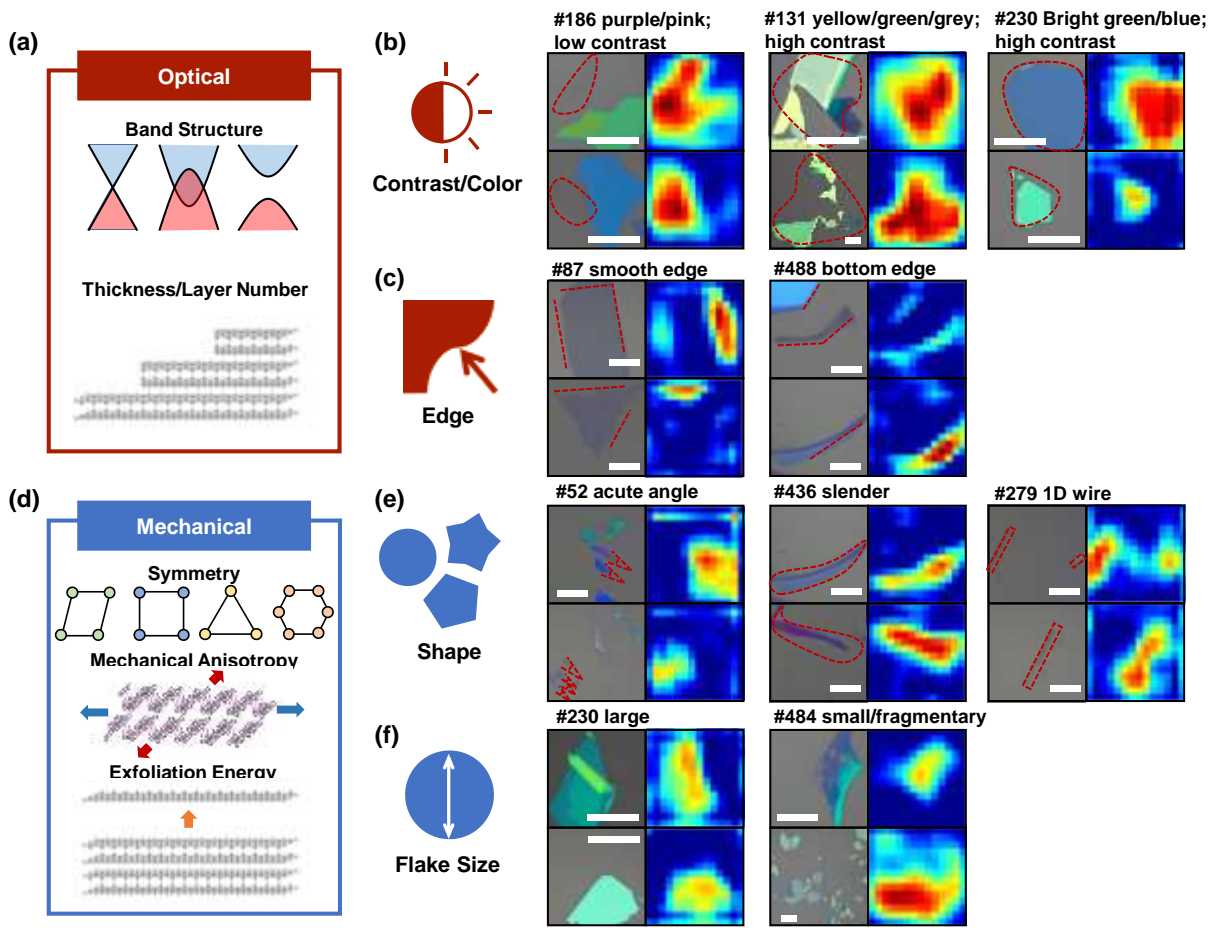

**Figure 3.** Deep graphical features captured by the 2DMOINet. (a) schematics of the physical properties such as the band structure and the thickness that determine the optical responses of the 2D flakes. (b) Contrast/color and (c) edge and typical feature maps in the Depth=5 layer of the 2DMOINet that are associated with the optical responses. (d) schematics of the physical properties such as the crystal symmetry, the mechanical anisotropy and the exfoliation energy that determine the mechanical responses of the 2D flakes. (e) flake shape and (f) flake size and typical feature maps in the Depth=5 layer of the 2DMOINet that are correlated to the mechanical properties of the materials. The high-activation regions in the feature maps are also indicated by red dashed curves in the corresponding OM images. Scale bars: 20 μm

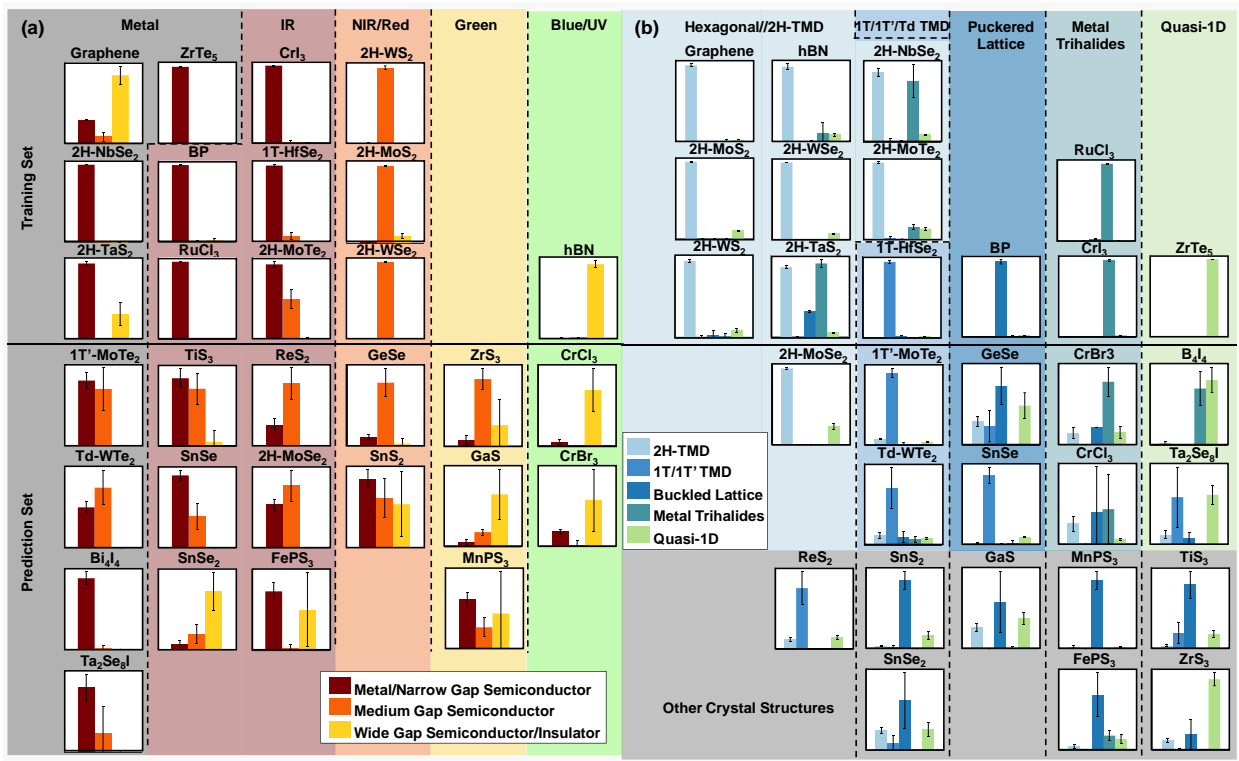

**Figure 4.** Prediction of physical properties through ensemble learning based on 16 independently trained 2DMOINets. (a) histograms of the projected values of the materials in the training set (top half) and in the prediction set (bottom half, unused when training the 2DMOINets) that predict the bandgap of the materials. (b) histograms of the projected values that predict the crystal structure of the materials. The materials are regrouped by the bandgaps and the crystal structures respectively in (a) and (b). The histograms and the error bars are the mean projected values and their standard deviations of the material property predictor constructed based on 16 independently trained 2DMOINet. The training set contains the 13 materials used for the training of the 2DMOINet, whereas the prediction set contains 17 additional materials that are unknown to the

2DMOINet during the training stage. The materials in the grey box in (b) do not belong to any of the crystal structure classes.

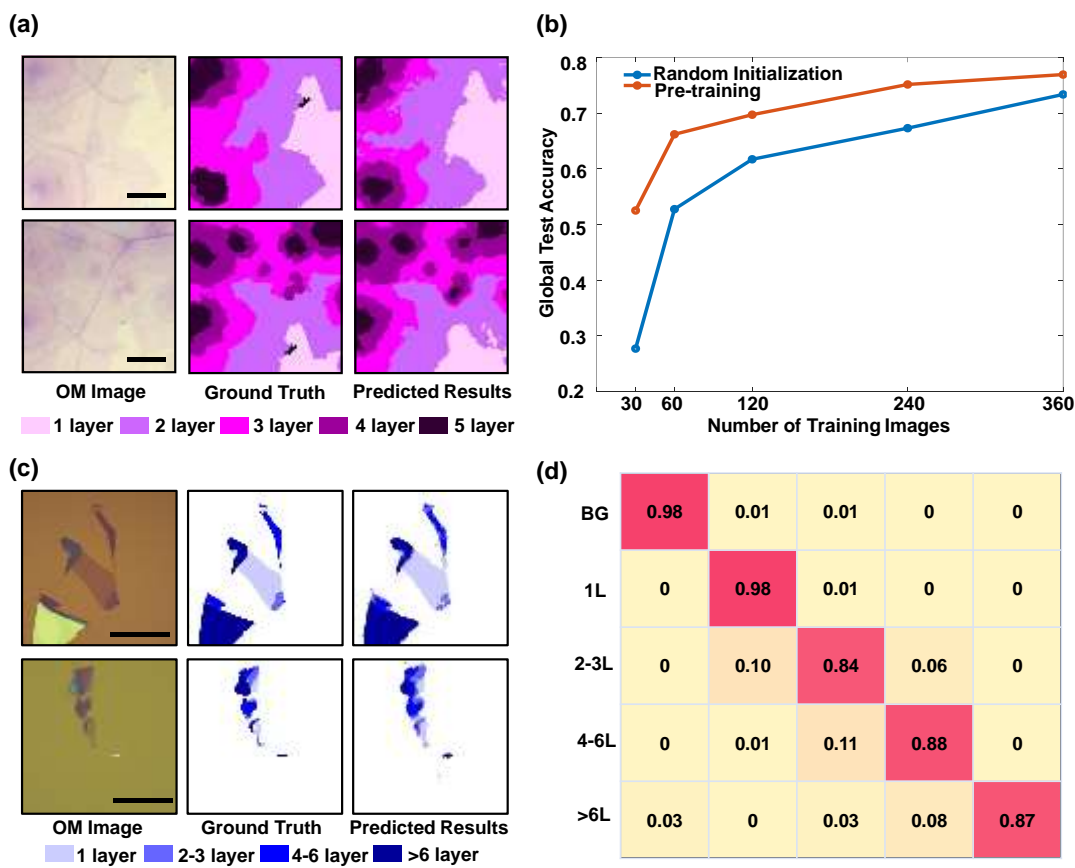

**Figure 5.** Transfer learning for CVD graphene and exfoliated Td-WTe$_2$. (a,c) Typical training results on the CVD graphene dataset (a) and on the exfoliated Td-WTe$_2$ dataset. The left column are OM images, the middle column are the ground truth label maps, and the right column are the label maps predicted by the re-trained 2DMOINet. Scale bars: 20 μm. (b) Global test accuracy as a function of the number of images (CVD graphene) in the training dataset for the pre-training

method (red) and the random initialization method (blue). (d) Pixel-level confusion matrix on the test dataset of the 2DMOINet trained for exfoliated Td-WTe$_2$. The mean class accuracy reaches 91%.

**Table 1.** Overall classification performance of the 2DMOINet. The global accuracy (by pixel), the mean accuracy (by class) and the mean IoU (by pixel) are all calculated on the test dataset. The experiment environment is: CPU: Intel(R) Core(TM) i7-8700K CPU @ 3.70GHz, 32.0GB RAM; GPU: NVIDIA GeForce GTX 1080 Ti, 11 GB GDDR5X.

| Global accuracy (by pixel) | Mean accuracy (by class) | Mean IoU (by pixel) | Training time | Frames per second (FPS) in test | |
|---|---|---|---|---|---|
| | | | | Use CPU | Use GPU |
| 0.9689 | 0.7978 | 0.5878 | 30 hrs 56mins 23s | 2.5 | 22.0 |

# Reference

1. Long J, Shelhamer E, Darrell T. Fully convolutional networks for semantic segmentation.  2015 IEEE Conference on Computer Vision and Pattern Recognition (CVPR); 2015 7-12 June 2015; 2015. p. 3431-3440.

2. Chen L, Papandreou G, Kokkinos I, Murphy K, Yuille AL. DeepLab: Semantic Image Segmentation with Deep Convolutional Nets, Atrous Convolution, and Fully Connected CRFs. *IEEE Transactions on Pattern Analysis and Machine Intelligence* 2018, **40**(4)**:** 834-848.

3. Ronneberger O, Fischer P, Brox T. U-Net: Convolutional Networks for Biomedical Image Segmentation. In: Navab N, Hornegger J, Wells WM, Frangi AF, editors. Medical Image Computing and Computer-Assisted Intervention – MICCAI 2015; 2015 2015//; Cham: Springer International Publishing; 2015. p. 234-241.

4. Badrinarayanan V, Kendall A, Cipolla R. SegNet: A Deep Convolutional Encoder-Decoder Architecture for Image Segmentation. *IEEE Transactions on Pattern Analysis and Machine Intelligence* 2017, **39**(12)**:** 2481-2495.

5. Geim AK, Novoselov KS. The rise of graphene. *Nature Materials* 2007, **6:** 183.

6. Ferrari AC, Bonaccorso F, Fal'ko V, Novoselov KS, Roche S, Bøggild P*, et al.* Science and technology roadmap for graphene, related two-dimensional crystals, and hybrid systems. *Nanoscale* 2015, **7**(11)**:** 4598-4810.

7. Nicolosi V, Chhowalla M, Kanatzidis MG, Strano MS, Coleman JN. Liquid Exfoliation of Layered Materials. *Science* 2013, **340**(6139)**:** 1226419.

8. Wang QH, Kalantar-Zadeh K, Kis A, Coleman JN, Strano MS. Electronics and optoelectronics of two-dimensional transition metal dichalcogenides. *Nature Nanotechnology* 2012, **7:** 699.

9. Tan C, Cao X, Wu X-J, He Q, Yang J, Zhang X*, et al.* Recent Advances in Ultrathin Two-Dimensional Nanomaterials. *Chemical Reviews* 2017, **117**(9)**:** 6225-6331.

10. Novoselov KS, Geim AK, Morozov SV, Jiang D, Zhang Y, Dubonos SV*, et al.* Electric Field Effect in Atomically Thin Carbon Films. *Science* 2004, **306**(5696)**:** 666-669.

11. Yi M, Shen Z. A review on mechanical exfoliation for the scalable production of graphene. *Journal of Materials Chemistry A* 2015, **3**(22)**:** 11700-11715.

12. Masubuchi S, Morimoto M, Morikawa S, Onodera M, Asakawa Y, Watanabe K, *et al.* Autonomous robotic searching and assembly of two-dimensional crystals to build van der Waals superlattices. *Nature Communications* 2018, **9**(1)**:** 1413.

13. Li H, Wu J, Huang X, Lu G, Yang J, Lu X, *et al.* Rapid and Reliable Thickness Identification of Two-Dimensional Nanosheets Using Optical Microscopy. *ACS Nano* 2013, **7**(11)**:** 10344-10353.

14. Lin X, Si Z, Fu W, Yang J, Guo S, Cao Y, *et al.* Intelligent identification of two-dimensional nanostructures by machine-learning optical microscopy. *Nano Research* 2018, **11**(12)**:** 6316-6324.

15. Masubuchi S, Machida T. Classifying optical microscope images of exfoliated graphene flakes by data-driven machine learning. *npj 2D Materials and Applications* 2019, **3**(1)**:** 4.

16. Ni ZH, Wang HM, Kasim J, Fan HM, Yu T, Wu YH, *et al.* Graphene Thickness Determination Using Reflection and Contrast Spectroscopy. *Nano Letters* 2007, **7**(9)**:** 2758-2763.

17. Nolen CM, Denina G, Teweldebrhan D, Bhanu B, Balandin AA. High-Throughput Large-Area Automated Identification and Quality Control of Graphene and Few-Layer Graphene Films. *ACS Nano* 2011, **5**(2)**:** 914-922.

18. Blake P, Hill EW, Castro Neto AH, Novoselov KS, Jiang D, Yang R, *et al.* Making graphene visible. *Applied Physics Letters* 2007, **91**(6)**:** 063124.

19. [cited]Available from: https://www.2dsemiconductors.com/; http://www.hqgraphene.com/

20. Sutskever I, Martens J, Dahl G, Hinton G. On the importance of initialization and momentum in deep learning. International conference on machine learning; 2013; 2013. p. 1139-1147.

21. Simonyan K, Zisserman A. Very deep convolutional networks for large-scale image recognition. *arXiv preprint arXiv:14091556* 2014.

22. Szegedy C, Wei L, Yangqing J, Sermanet P, Reed S, Anguelov D, *et al.* Going deeper with convolutions. 2015 IEEE Conference on Computer Vision and Pattern Recognition (CVPR); 2015 7-12 June 2015; 2015. p. 1-9.

23. Boykov YY, Jolly M. Interactive graph cuts for optimal boundary & region segmentation of objects in N-D images. Proceedings Eighth IEEE International Conference on Computer Vision. ICCV 2001; 2001 7-14 July 2001; 2001. p. 105-112 vol.101.

Supplementary Information

# Deep-Learning-Enabled Fast Optical Identification and Characterization of Two-Dimensional Materials

**This document contains:**

**Tables S1-S6**

**Figures S1-S54**

**Table S1.** Detailed information of the SegNet based on the VGG16 network.

|  | Encoder | Decoder |
|---|---|---|
|  | Input | Output |
|  | 224 by 224 RGB image | Softmax |
| **Depth=1** | Conv 3*3*64 | Conv 3*3*64 |
|  | Conv 3*3*64 | Conv 3*3*64 |
|  | Maxpooling | Upsampling 2*2 |
| **Depth=2** | Conv 3*3*128 | Conv 3*3*128 |
|  | Conv 3*3*128 | Conv 3*3*128 |
|  | Maxpooling 2*2 | Upsampling 2*2 |
| **Depth=3** | Conv 3*3*256 | Conv 3*3*256 |
|  | Conv 3*3*256 | Conv 3*3*256 |
|  | Conv 3*3*256 | Conv 3*3*256 |
|  | Maxpooling 2*2 | Upsampling 2*2 |
| **Depth=4** | Conv 3*3*512 | Conv 3*3*512 |
|  | Conv 3*3*512 | Conv 3*3*512 |
|  | Conv 3*3*512 | Conv 3*3*512 |
|  | Maxpooling 2*2 | Upsampling 2*2 |
| **Depth=5** | Conv 3*3*512 | Conv 3*3*512 |
|  | Conv 3*3*512 | Conv 3*3*512 |
|  | Conv 3*3*512 | Conv 3*3*512 |
|  | Maxpooling 2*2 | Upsampling 2*2 |

**Table S2.** Classification performance of SegNet under different depth (each depth has 2 conv layer, the dimension of convolutional kernel is 64).

|  | Global Accuracy (by pixel) | Mean Accuracy (by class) | Mean IoU (by pixel) |
|---|---|---|---|
| **Depth=1** | 0.8925 | 0.4291 | 0.2067 |
| **Depth=2** | 0.9016 | 0.4751 | 0.2322 |
| **Depth=3** | 0.9164 | 0.5558 | 0.2757 |
| **Depth=4** | 0.9232 | 0.6225 | 0.3363 |
| **Depth=5** | **0.9414** | **0.7063** | **0.4219** |
| **Depth=6** | 0.9145 | 0.5875 | 0.3152 |

**Table S3.** The overall performance of SegNet under different data augmentation method.

| Basic | Color | Rotation | Size of training dataset | Epoch | Global accuracy (by pixel) | Mean accuracy (by class) | Mean IoU (by pixel) |
|---|---|---|---|---|---|---|---|
|  |  |  | 3,825 | 20 | 0.9250 | 0.5865 | 0.3287 |
|  | √ |  | 49,725 | 20 | 0.9594 | 0.6172 | 0.4710 |
| √ |  |  | 22,950 | 20 | 0.9611 | **0.7798** | 0.5347 |
| √ | √ |  | **68,850** | 20 | **0.9681** | 0.7435 | **0.5570** |
|  |  | √ | 3,825* | 250 | **0.9670** | **0.8046** | **0.5801** |

**Note:** (*) The rotation transformation is applied randomly in each epoch during the training. As a result, although the number of images in the training dataset does not increase, the SegNet could "see" much more variations of these images during the training stage.

**Table S4.** The overall performance of 16 SegNets trained separately. Networks #1, #2 are trained with the basic data augmentation; networks #3-9 are trained with random rotation data augmentation and 100 additional background-only images; networks #10-16 are trained with random rotation data augmentation and no background-only images. Network #5 has the best performance. Most of the analysis in this work is made based on Network #5 if not stated specifically.

| # | Training dataset | Data augmentation | Global accuracy (by pixel) | Mean accuracy (by class) | Mean IoU (by pixel) |
|---|---|---|---|---|---|
| 1 | No BG | Basic | 0.9611 | 0.7798 | 0.5347 |
| 2 | No BG | Basic | 0.9612 | 0.7786 | 0.5303 |
| 3 | +100BG | Random rotation | 0.9678 | 0.7960 | 0.5806 |
| 4 | +100BG | Random rotation | 0.9648 | 0.7792 | 0.5643 |
| **5** | **+100BG** | **Random rotation** | **0.9689** | **0.7978** | **0.5878** |
| 6 | +100BG | Random rotation | 0.9661 | 0.8026 | 0.5771 |
| 7 | +100BG | Random rotation | 0.9675 | 0.7895 | 0.5812 |
| 8 | +100BG | Random rotation | 0.9676 | 0.7914 | 0.5819 |
| 9 | +100BG | Random rotation | 0.9672 | 0.7865 | 0.5765 |
| 10 | No BG | Random rotation | 0.9679 | 0.8020 | 0.5856 |
| 11 | No BG | Random rotation | 0.9681 | 0.7967 | 0.5837 |
| 12 | No BG | Random rotation | 0.9670 | 0.8046 | 0.5801 |
| 13 | No BG | Random rotation | 0.9670 | 0.7880 | 0.5755 |
| 14 | No BG | Random rotation | 0.9661 | 0.7837 | 0.5683 |
| 15 | No BG | Random rotation | 0.9683 | 0.7989 | 0.5831 |
| 16 | No BG | Random rotation | 0.9662 | 0.8032 | 0.5750 |

**Table S5.** A summary of the physical properties of the 2D materials (bulk) used in this study [24, 25, 26, 27, 28, 29, 30, 31, 32, 33, 34, 35, 36, 37, 38, 39, 40]. The values for the optical bandgap the direct bandgap, or the lowest peak energy in the optical absorption spectra.

| Material | Crystal System | Point Group | Space Group | Bandgap (Optical) | Exfoliation Energy |
|---|---|---|---|---|---|
| graphite | Hexagonal | $D_{6h}$ | P6/mmm | 0 eV | 70.36 meV |
| hBN | Hexagonal | $D_{6h}$ | $P6_3$/mmc | ~6 eV | 71.34 meV |
| 2H-$MoS_2$ | Hexagonal | $D_{6h}$ | $P6_3$/mmc | 1.8 eV | 76.99 meV |
| 2H-$WS_2$ | Hexagonal | $D_{6h}$ | $P6_3$/mmc | 2.1 eV | 76.27 meV |
| 2H-$WSe_2$ | Hexagonal | $D_{6h}$ | $P6_3$/mmc | 1.7 eV | 79.63 meV |
| 2H-$MoTe_2$ | Hexagonal | $D_{6h}$ | $P6_3$/mmc | 1.1 eV | 90.98 meV |
| 2H-$TaS_2$ | Hexagonal | $D_{6h}$ | $P6_3$/mmc | 0 eV | 87.15 meV |
| 2H-$NbSe_2$ | Hexagonal | $D_{6h}$ | $P6_3$/mmc | 0 eV | 98.28 meV |
| 1T-$HfSe_2$ | Trigonal | $D_{3d}$ | $P\bar{3}m1$ | 1.1 eV | 92.05 meV |
| BP | Orthorhombic | $D_{2h}$ | Cmce | 0.35 eV | 111.62 meV |
| $CrI_3$ | Monoclinic | $C_{2h}$ | C2/m | 1.2 eV | - |
| $RuCl_3$ | Monoclinic | $C_{2h}$ | C2/m | 0.3 eV | - |
| $ZrTe_5$ | Orthorhombic | $D_{2h}$ | Cmcm | 0 eV | 90.00 meV |
| 2H-$MoSe_2$ | Hexagonal | $D_{6h}$ | $P6_3/mmc$ | 1.5 eV | 80.24 meV |
| 1T'-$MoTe_2$ | Monoclinic | $C_{2h}$ | $P2_1/m$ | 0 eV | 86.87 meV |
| Td-$WTe_2$ | Orthorhombic | $C_{2v}$ | $Pmn2_1$ | 0 eV | 84.90 meV |
| GeSe | Orthorhombic | $D_{2h}$ | Pnma | 2.1 eV | - |
| SnSe | Orthorhombic | $D_{2h}$ | Pnma | 1.3 eV | 152.65 meV |
| $CrBr_3$ | Monoclinic | $C_{2h}$ | C2/m | 3.2 eV | - |
| $CrCl_3$ | Monoclinic | $C_{2h}$ | C2/m | 3.1 eV | 69.52 meV |
| $Bi_4I_4$ | Monoclinic | $C_{2h}$ | C2/m | 0.04 eV | 77.69 meV |
| $Ta_2Se_8I$ | Tetragonal | $D_4$ | I422 | 0 eV | - |
| GaS | Hexagonal | $D_{6h}$ | $P6_3/mmc$ | 2.5 eV | 56.22 meV |
| $ReS_2$ | Triclinic | $C_i$ | $P\bar{1}$ | 1.4 eV | 71.00 meV |
| $SnS_2$ | Trigonal | $D_{3d}$ | $P\bar{3}m1$ | 2.4 eV | 83.28 meV |
| $SnSe_2$ | Trigonal | $D_{3d}$ | $P\bar{3}m1$ | 1.6 eV | 93.47 meV |
| $MnPS_3$ | Monoclinic | $C_{2h}$ | C2/m | 2.8 eV | - |
| $FePS_3$ | Monoclinic | $C_{2h}$ | C2/m | 1.5 eV | - |
| $TiS_3$ | Monoclinic | $C_{2h}$ | $P2_1/m$ | 1.1 eV | 54.49 meV |
| $ZrS_3$ | Monoclinic | $C_{2h}$ | $P2_1/m$ | 2.5 eV | 55.49 meV |

**Table S6.** A summary of selected channels and their interpretations in the Depth=5 encoder layer of the trained SegNet.

| Channel # | Cases | Contrast/Color | Edge | Shape | Flake Size |
|---|---|---|---|---|---|
| 52 | | | | Acute Angle | |
| 53 | - | Purple/pink/red Low contrast | Right and bottom edges | | |
| 87 | - | Purple Low contrast | Straight/smooth edges | | |
| 131 | Case 1 | Yellow/green/gray | | | |
| | Case 2 | Non-uniform | | | |
| 132 | Case 1 | | | | Small/Fragmentary |
| | Case 2 | Non-uniform | | | |
| 143 | - | | | Small flakes next to large flakes | Fragmentary |
| 156 | - | | | Empty frame perimeters | |
| 179 | - | Purple | Left and bottom edges | | |
| 186 | - | Pink | | | Large |
| 230 | - | Bright green/blue High contrast | | | Large |
| 232 | | Green/gray | | | |
| 279 | | Purple | | 1D wires | |
| 283 | Case 1 | Purple | | 1D wires | |

|     | Case 2 | Non-uniform |  |  |  |
| --- | --- | --- | --- | --- | --- |
| 312 | Case 1 | Purple/pink |  | Slender |  |
|     | Case 2 | Purple/pink |  |  | Fragmentary |
| 322 | - | Purple/pink | Straight left edges |  |  |
| 402 | - |  |  | Empty frame perimeters |  |
| 425 | Case 1 |  |  | Empty frame perimeters |  |
|     | Case 2 |  |  | 1D wires |  |
| 436 | Case 1 | Purple/pink/blue |  | Slender |  |
|     | Case 2 | Purple/pink/blue |  |  | Fragmentary |
| 447 | - | Pink/red |  |  |  |
| 484 | Case 1 | Yellow |  |  |  |
|     | Case 2 | Green/pale green/grey |  |  | Fragmentary |
| 488 | Case 1 | Purple/pink/dark blue | Straight bottom edges |  |  |
|     | Case 2 | Purple/pink/dark blue |  | 1D wires |  |

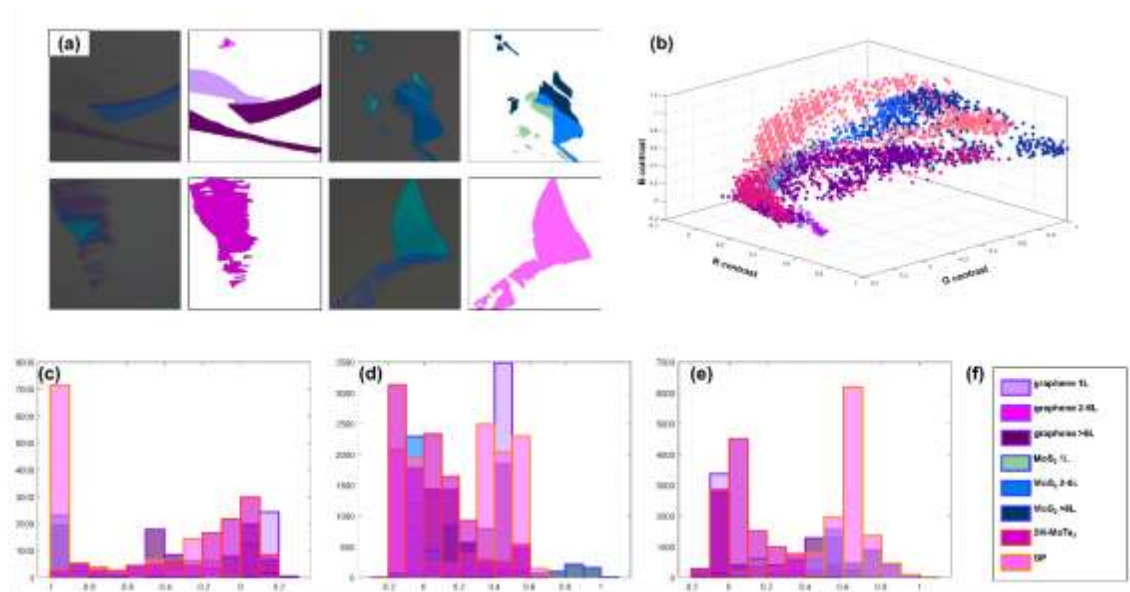

**Figure S1.** Optical contrast analysis of 4 different 2D materials, including graphene, $MoS_2$, 2H-$MoTe_2$, and BP. (a) typical OM images and the corresponding ground truth label maps. (b) RGB distribution of the images in (a). (c) Distribution of R values, (d) distributions of G values, and (e) distribution of B values of the images in (a). (f) color scales.

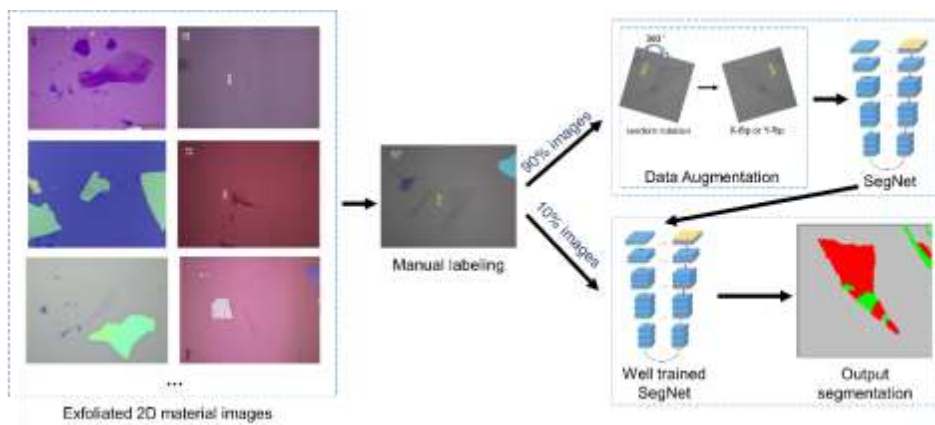

**Figure S2.** Schematic of the data generation procedure.

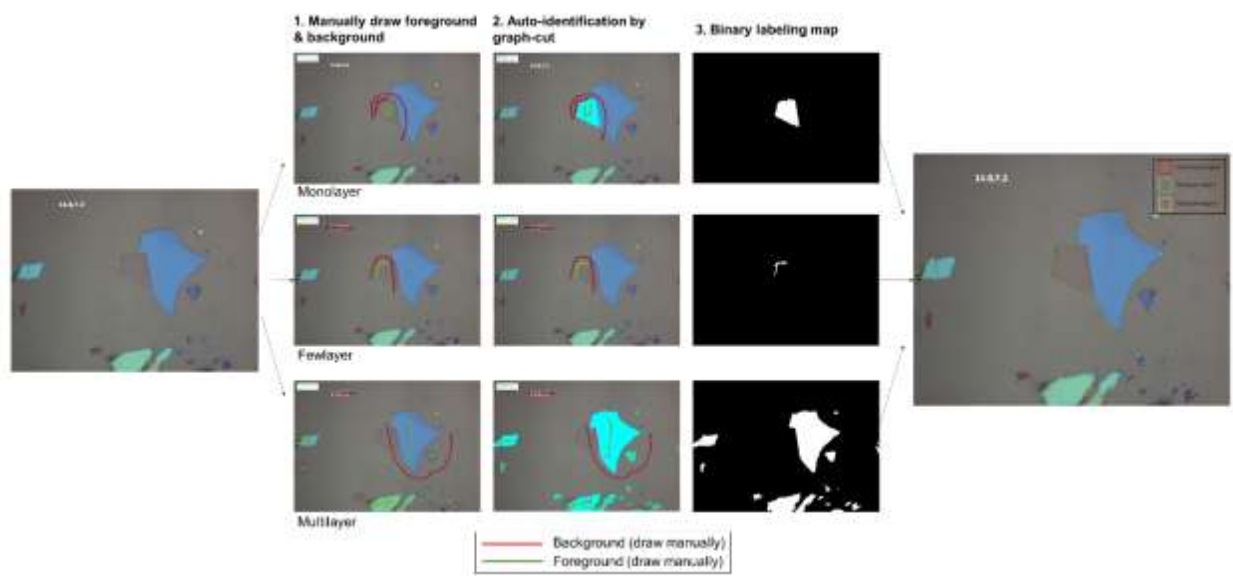

**Figure S3.** Schematic of the semi-automatic labeling procedure.

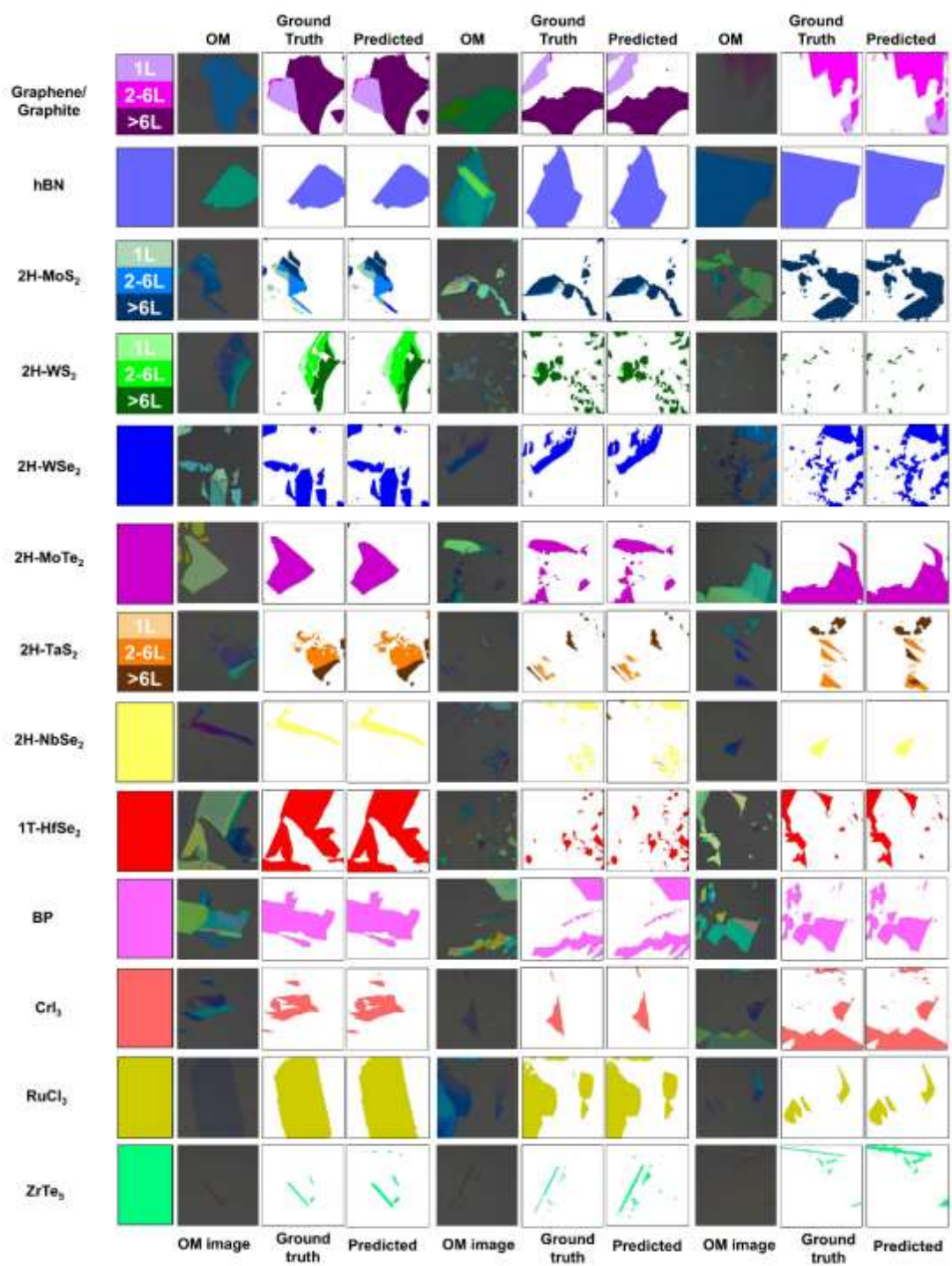

**Figure S4.** Additional results predicted by the SegNet.

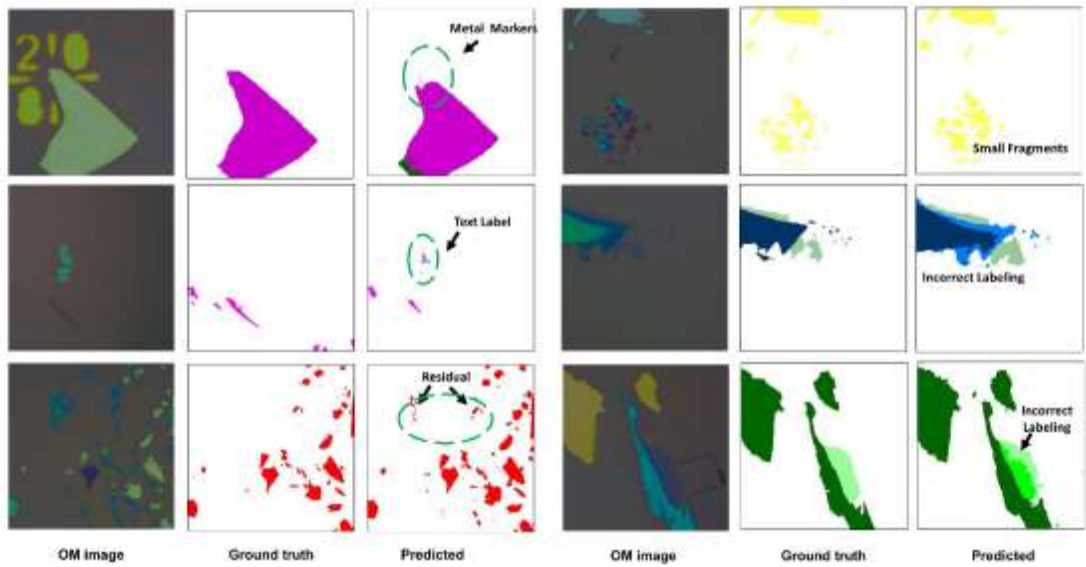

**Figure S5.** Examples of misclassified images.

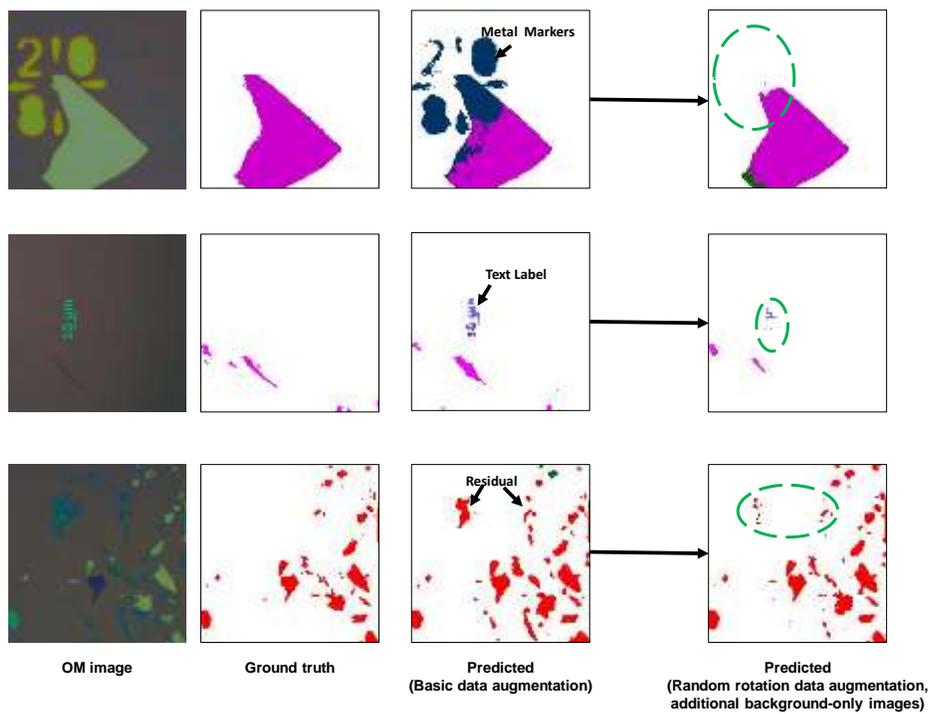

**Figure S6**. Comparison of predicted label maps by the SegNet trained with the basic data augmentation (third column) and the random data augmentation with additional background-only images (fourth column).

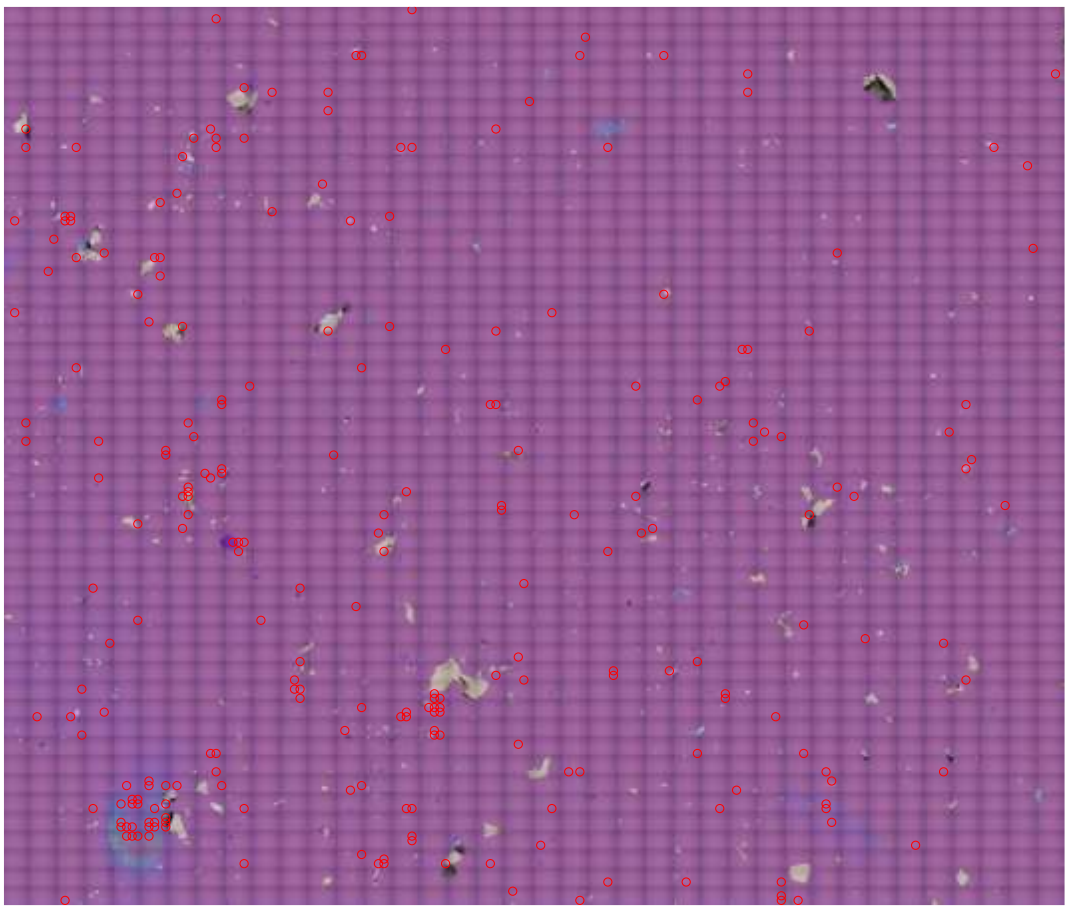

**Figure S7.** An overview OM image of the exfoliated graphene sample used to test the automatic 2D-material-searching system. The flakes of interest (monolayer and fewlayer segments with >100 pixels) are outlined in red.

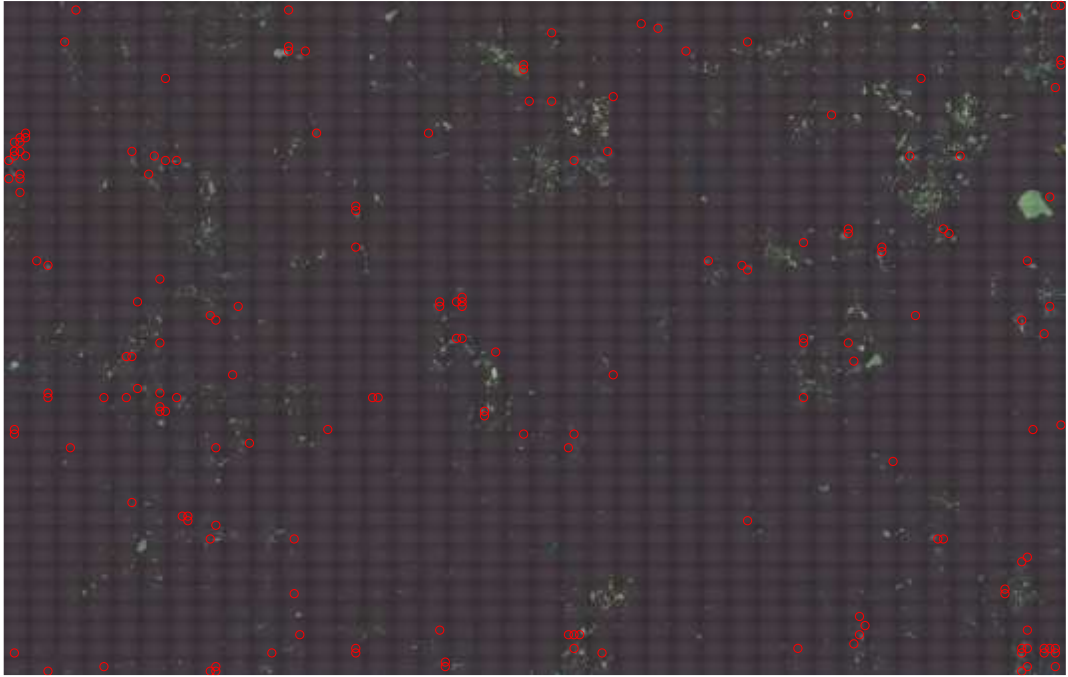

**Figure S8.** An overview OM image of the exfoliated MoS$_2$ sample used to test the automatic 2D-material-searching system. The flakes of interest (monolayer and fewlayer segments with >100 pixels) are outlined in red.

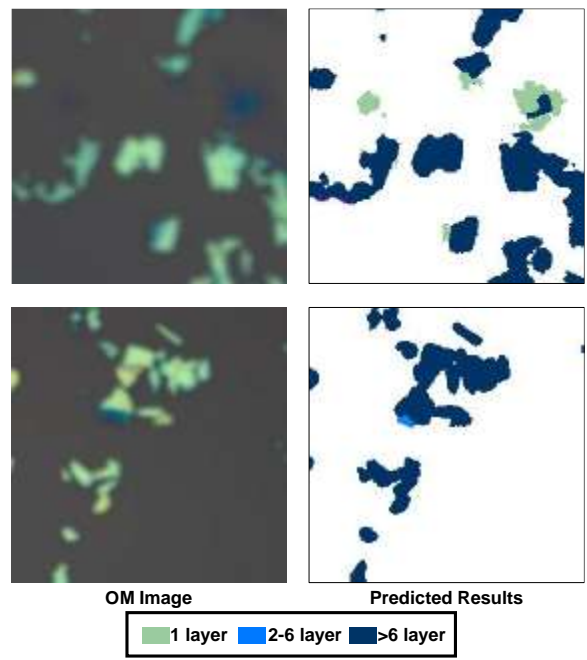

**Figure S9.** Examples of out-of-focus OM images and their corresponding predicted label maps taken on the exfoliated $MoS_2$ sample.

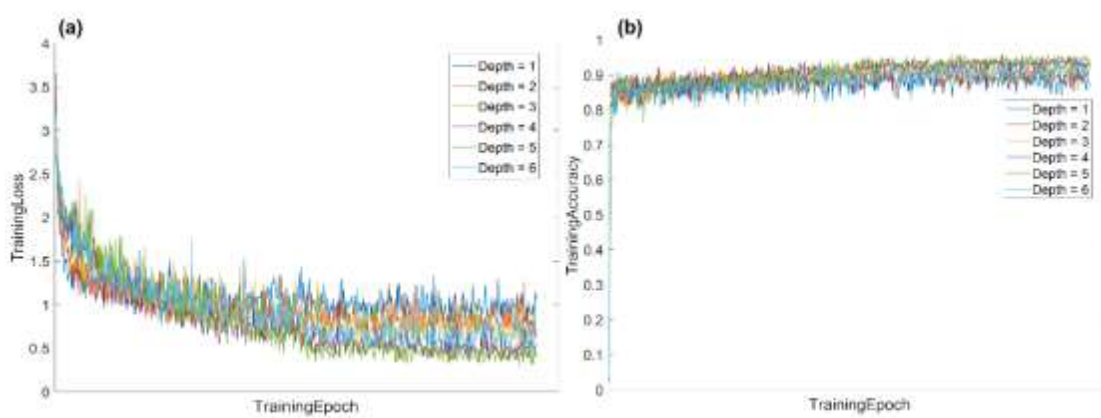

**Figure S10.** The trend of training loss and accuracy under different depth of network during training process.

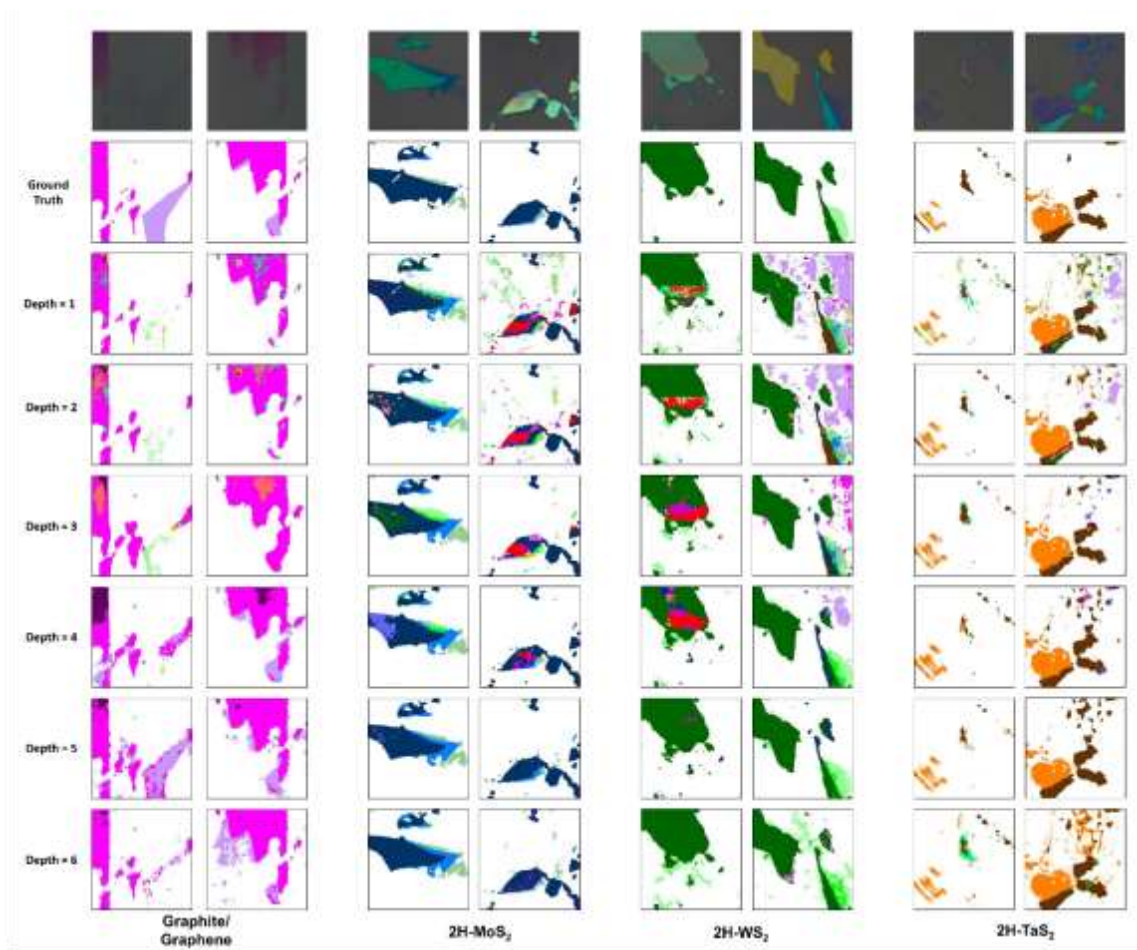

**Figure S11.** Semantic segmentation performance under different depth of network for Graphite/Graphene, 2H-MoS$_2$, 2H-WS$_2$ and 2H-TaS$_2$.

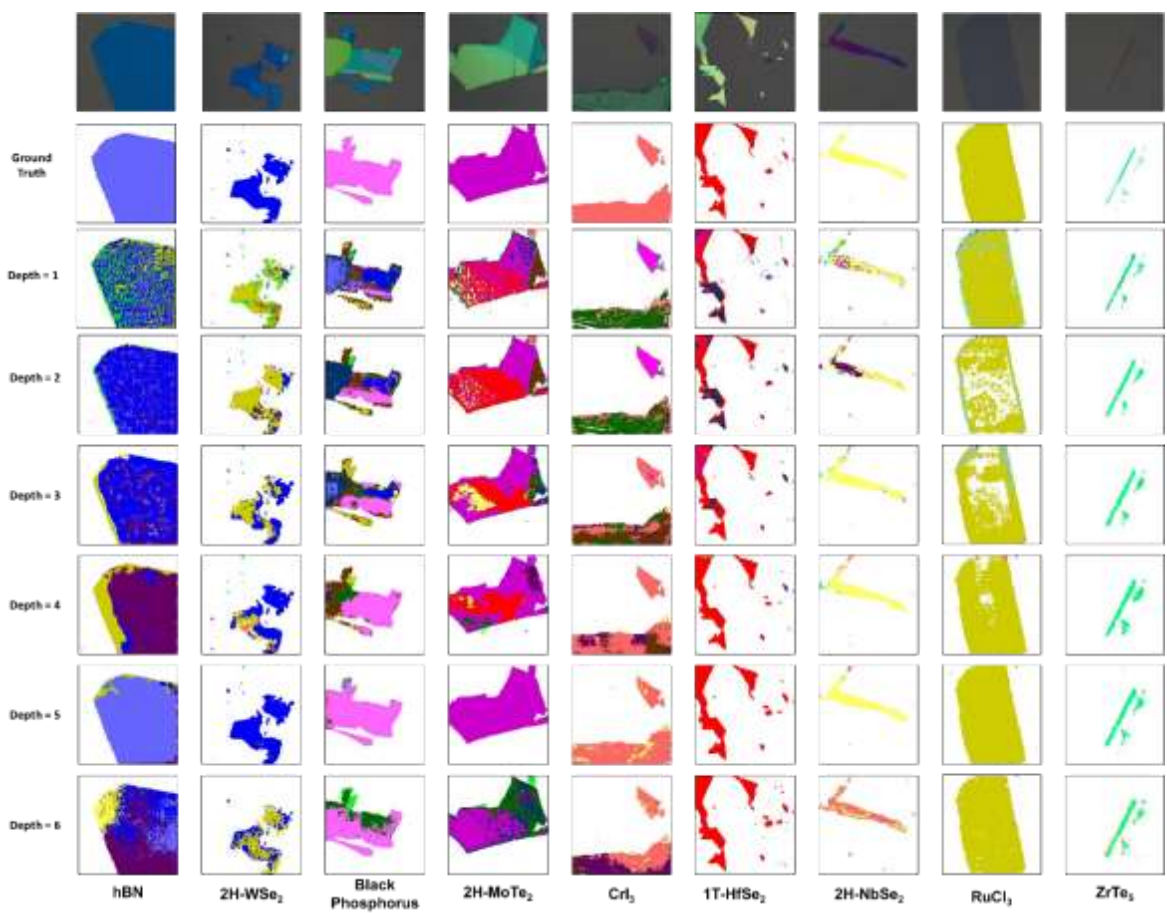

**Figure S12.** Semantic segmentation performance under different depth of network for hBN, 2H-WSe$_2$, Black Phosphorus, 2H-MoTe$_2$, CrI$_3$, 1T-HfSe$_2$, 2H-NbSe$_2$, RuCl$_3$ and ZrTe$_5$.

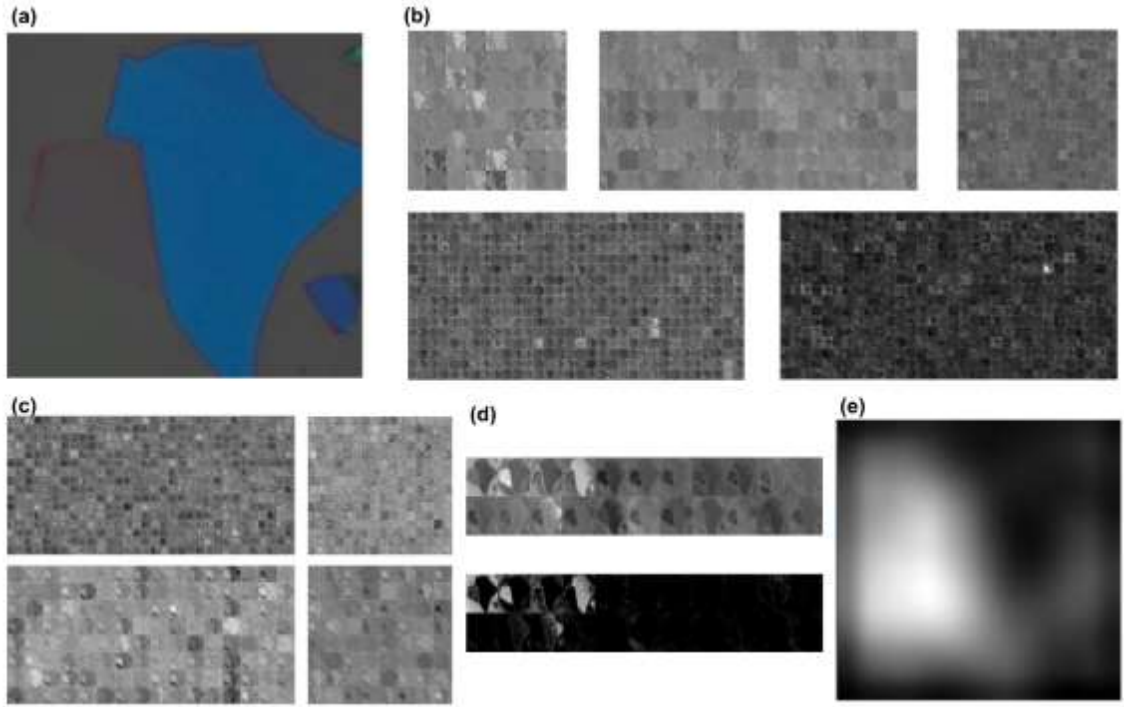

**Figure S13.** Feature maps in the trained SegNet. (a) the input OM image. (b) Depth=1 to 5 encoder layers. (c) Depth = 5 to 2 decoder layers. (d) The last decoder layers (conv. and ReLU layer). (e) Channel # 186 feature map of the Depth=5 encoder layer.

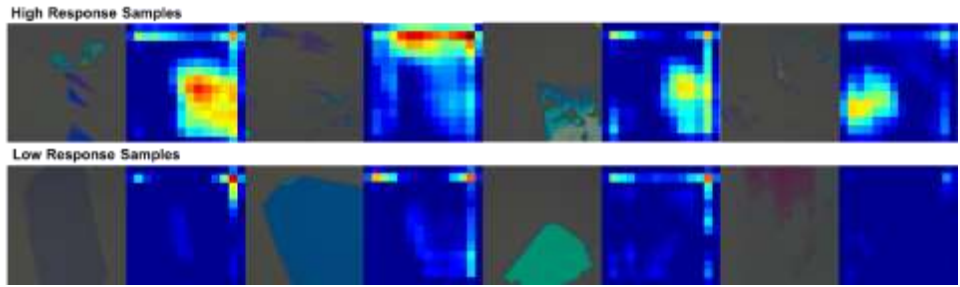

**Figure S14.** Represented optical images and their corresponding feature maps of Channel #52 of the Depth=5 encoder layer.

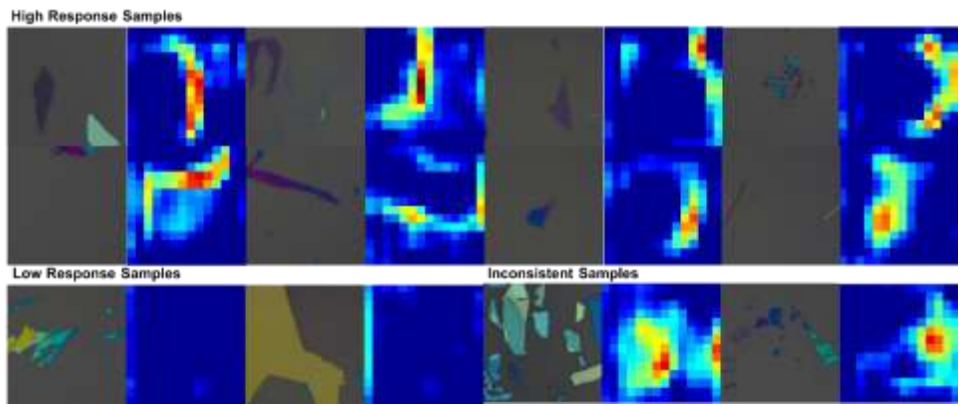

**Figure S15.** Represented optical images and their corresponding feature maps of Channel #53 of the Depth=5 encoder layer.

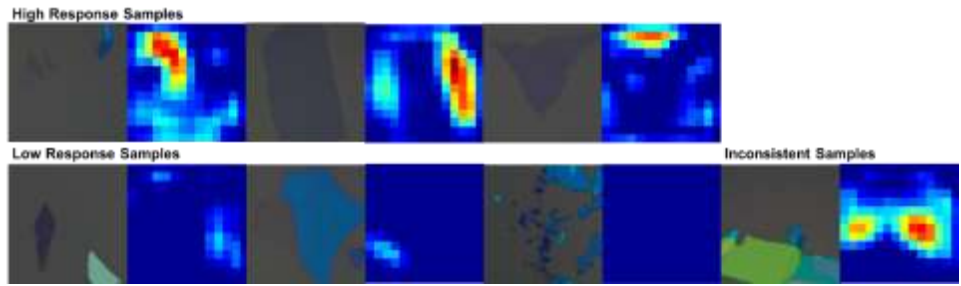

**Figure S16.** Represented optical images and their corresponding feature maps of Channel #87 of the Depth=5 encoder layer.

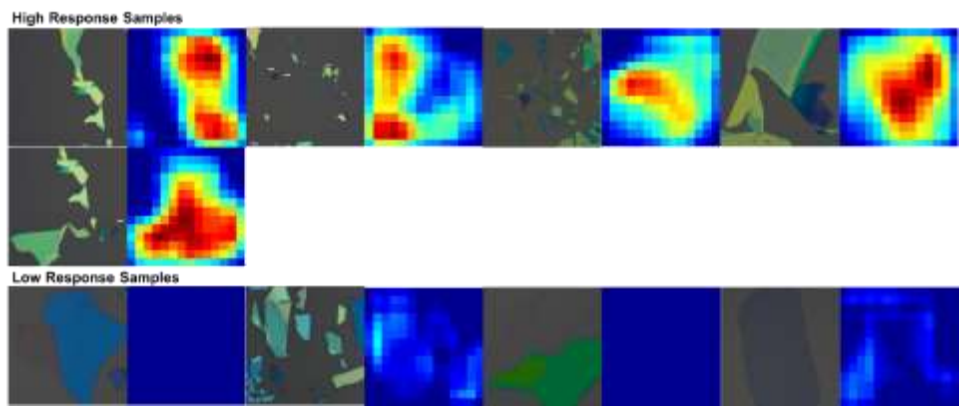

**Figure S17.** Represented optical images and their corresponding feature maps of Channel #131 of the Depth=5 encoder layer.

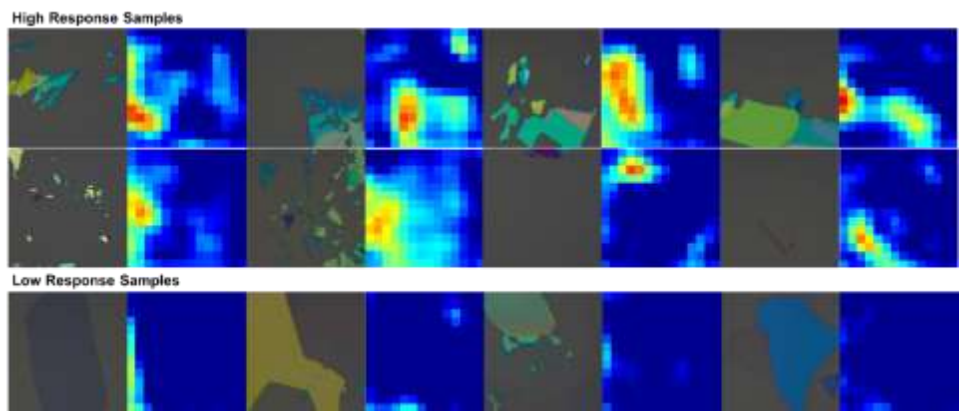

**Figure S18.** Represented optical images and their corresponding feature maps of Channel #132 of the Depth=5 encoder layer.

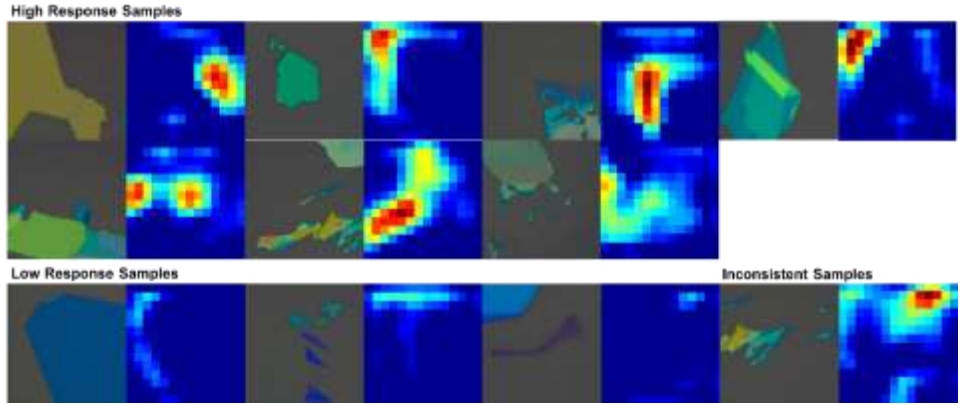

**Figure S19.** Represented optical images and their corresponding feature maps of Channel #143 of the Depth=5 encoder layer.

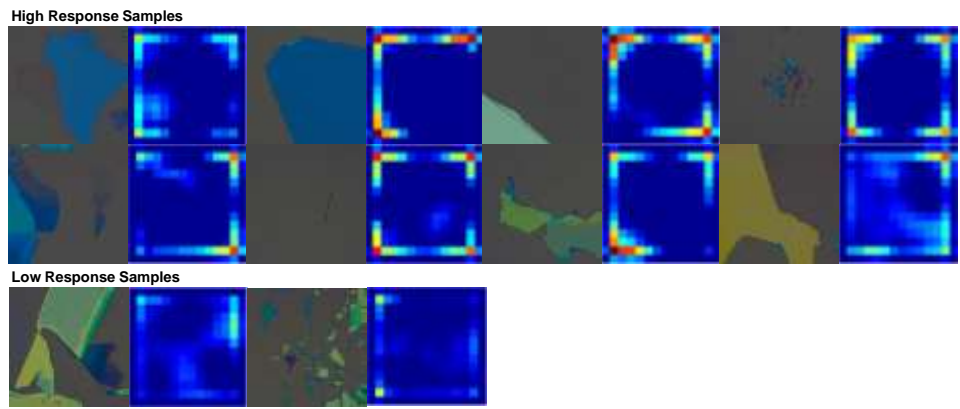

**Figure S20.** Represented optical images and their corresponding feature maps of Channel #156 of the Depth=5 encoder layer.

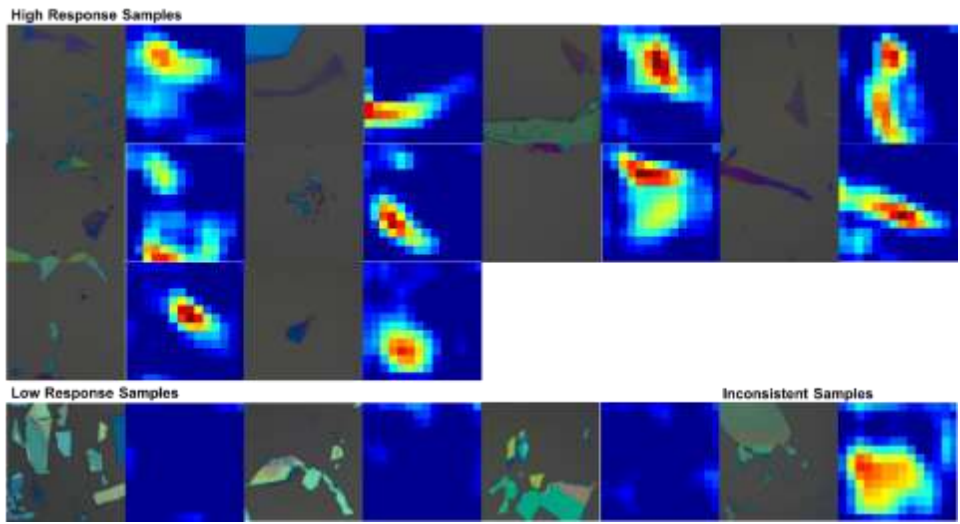

**Figure S21.** Represented optical images and their corresponding feature maps of Channel #179 of the Depth=5 encoder layer.

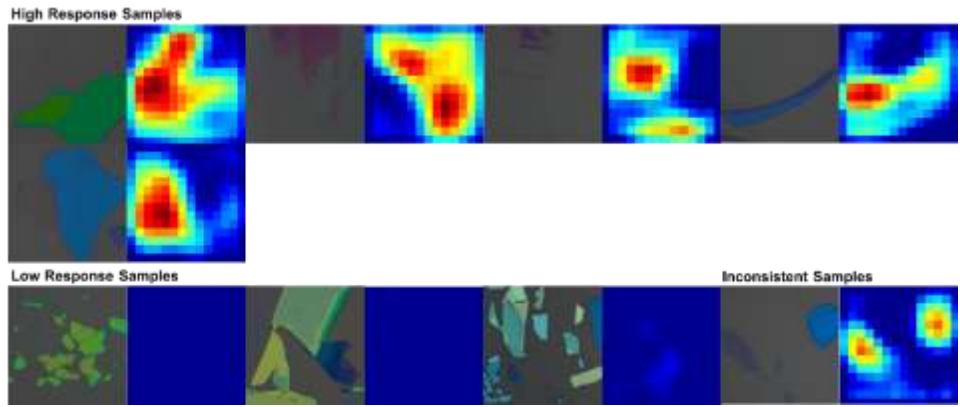

**Figure S22.** Represented optical images and their corresponding feature maps of Channel #186 of the Depth=5 encoder layer.

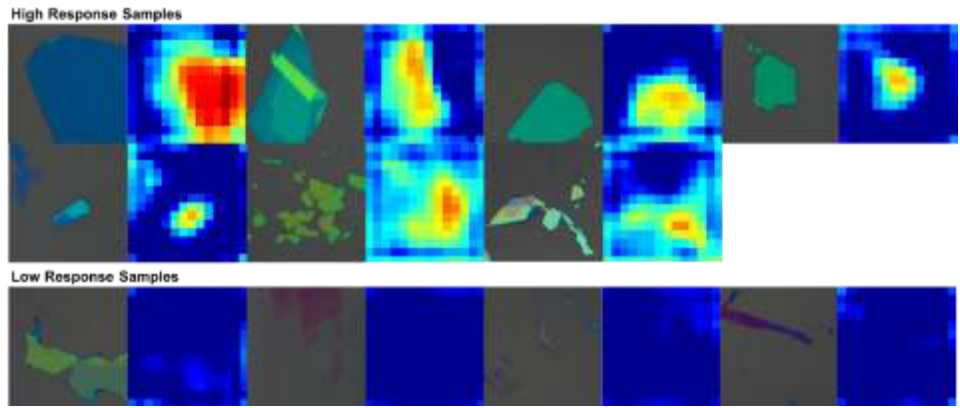

**Figure S23.** Represented optical images and their corresponding feature maps of Channel #230 of the Depth=5 encoder layer.

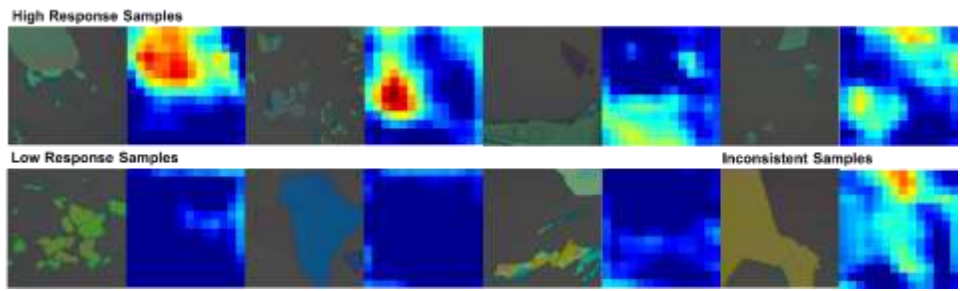

**Figure S24.** Represented optical images and their corresponding feature maps of Channel #232 of the Depth=5 encoder layer.

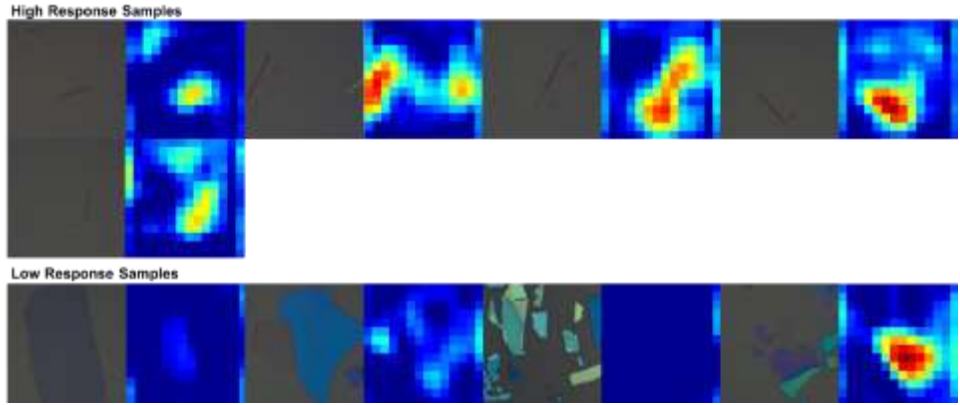

**Figure S25.** Represented optical images and their corresponding feature maps of Channel #279 of the Depth=5 encoder layer.

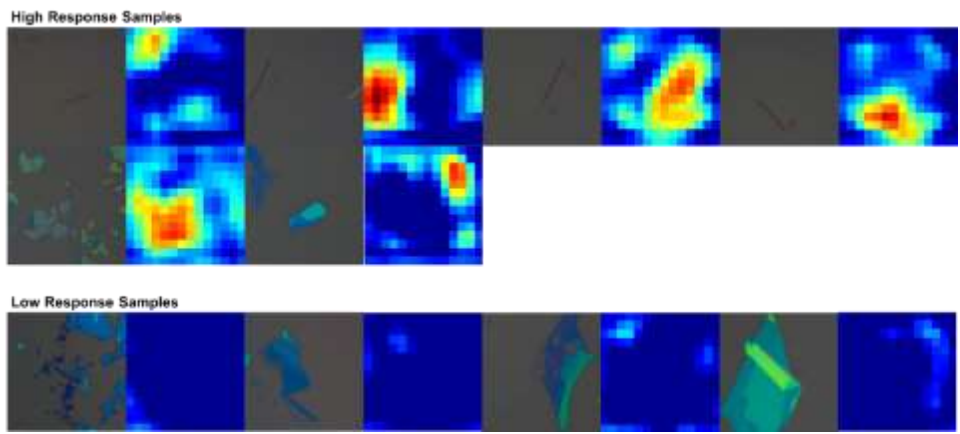

**Figure S26.** Represented optical images and their corresponding feature maps of Channel #283 of the Depth=5 encoder layer.

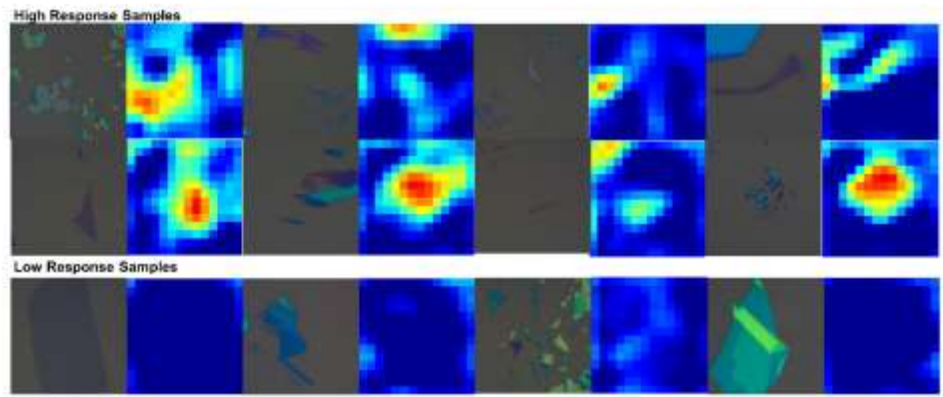

**Figure S27.** Represented optical images and their corresponding feature maps of Channel #312 of the Depth=5 encoder layer.

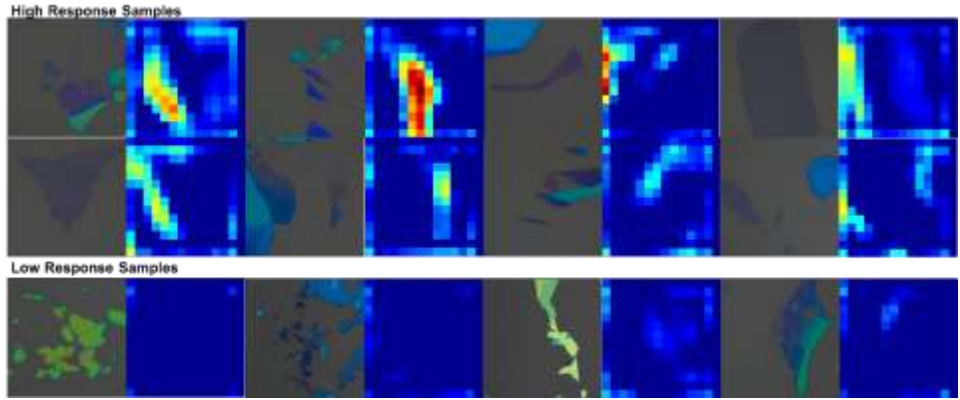

**Figure S28.** Represented optical images and their corresponding feature maps of Channel #322 of the Depth=5 encoder layer.

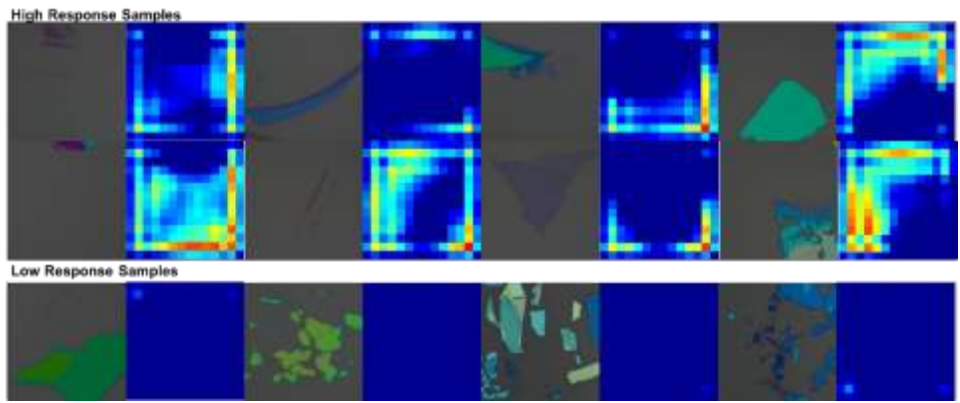

**Figure S29.** Represented optical images and their corresponding feature maps of Channel #402 of the Depth=5 encoder layer.

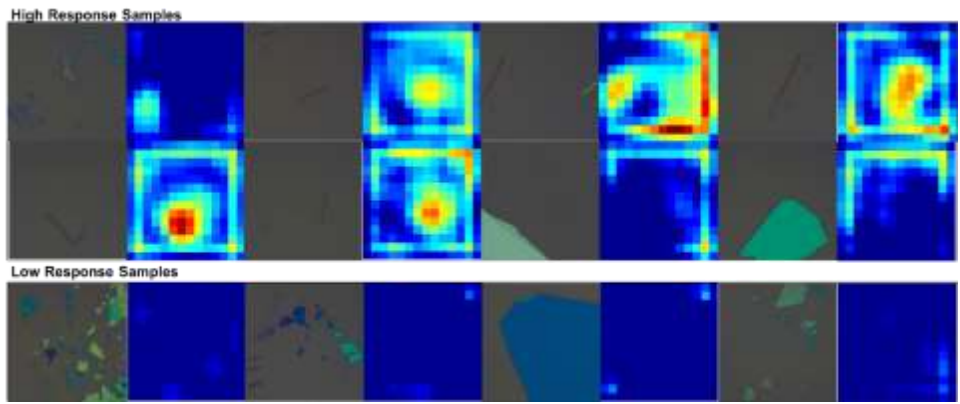

**Figure S30.** Represented optical images and their corresponding feature maps of Channel #425 of the Depth=5 encoder layer.

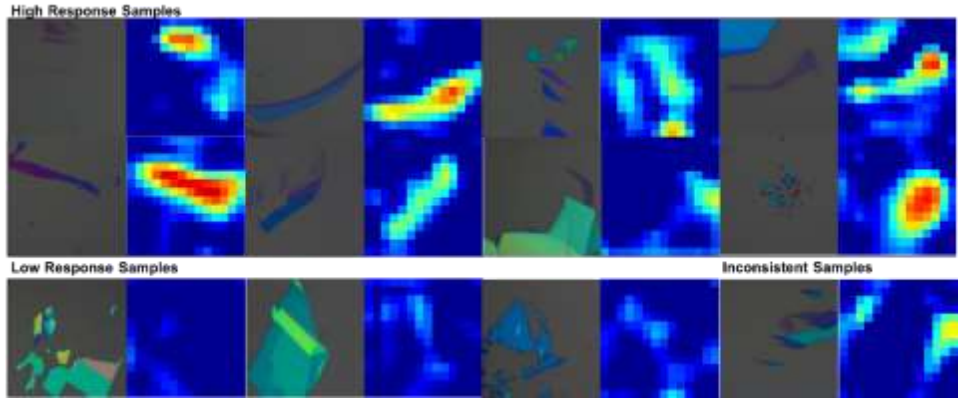

**Figure S31.** Represented optical images and their corresponding feature maps of Channel #436 of the Depth=5 encoder layer.

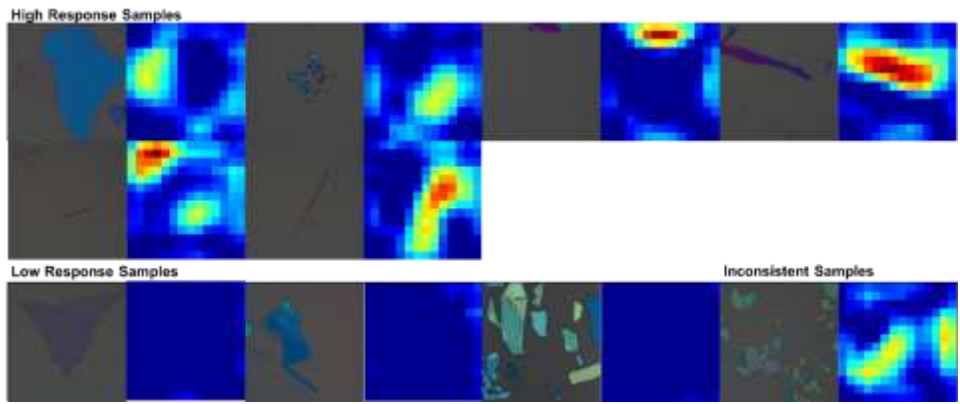

**Figure S32.** Represented optical images and their corresponding feature maps of Channel #447 of the Depth=5 encoder layer.

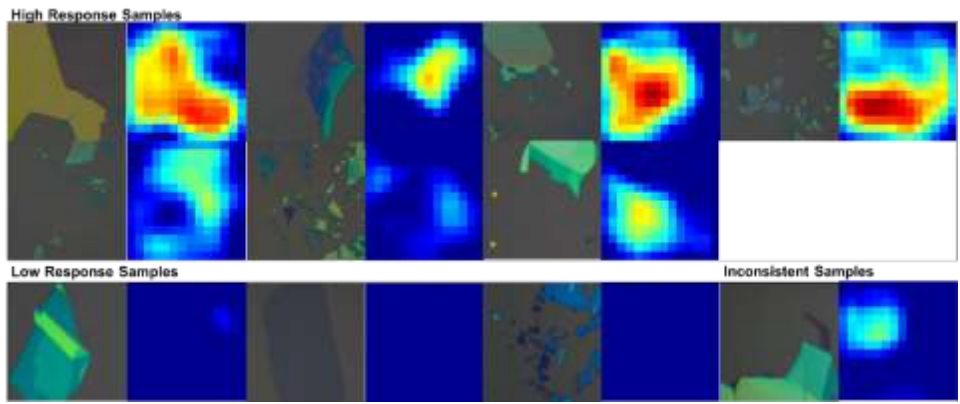

**Figure S33.** Represented optical images and their corresponding feature maps of Channel #484 of the Depth=5 encoder layer.

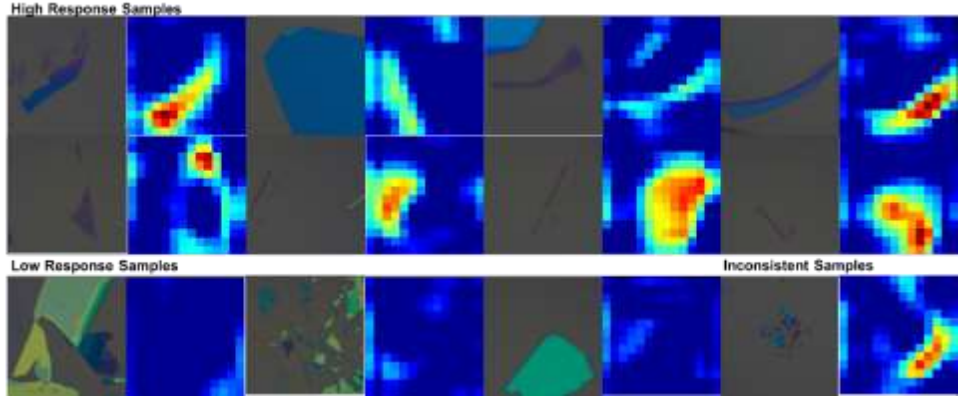

**Figure S34.** Represented optical images and their corresponding feature maps of Channel #488 of the Depth=5 encoder layer.

**Figure S35.** Extended confusion matrix for network #1 in Table S4.

**Figure S36.** Extended confusion matrix for network #2 in Table S4.

**Figure S37.** Extended confusion matrix for network #3 in Table S4.

**Figure S38.** Extended confusion matrix for network #4 in Table S4.

**Figure S39.** Extended confusion matrix for network #5 in Table S4.

**Figure S40.** Extended confusion matrix for network #6 in Table S4.

**Figure S41.** Extended confusion matrix for network #7 in Table S4.

**Figure S42.** Extended confusion matrix for network #8 in Table S4.

**Figure S43.** Extended confusion matrix for network #9 in Table S4.

**Figure S44.** Extended confusion matrix for network #10 in Table S4.

**Figure S45.** Extended confusion matrix for network #11 in Table S4.

**Figure S46.** Extended confusion matrix for network #12 in Table S4.

**Figure S47.** Extended confusion matrix for network #13 in Table S4.

**Figure S48.** Extended confusion matrix for network #14 in Table S4.

**Figure S49.** Extended confusion matrix for network #15 in Table S4.

**Figure S50.** Extended confusion matrix for network #16 in Table S4.

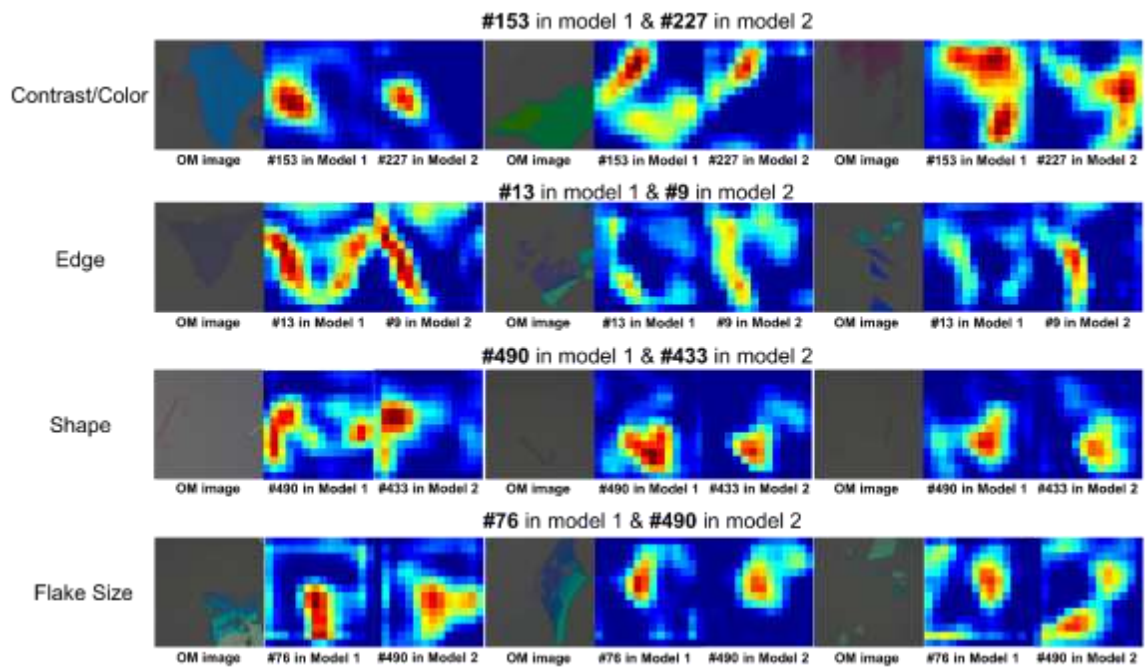

**Figure S51.** Depth=5 feature maps extracted from two independently trained SegNet that show the same activation behaviors on the same OM images.

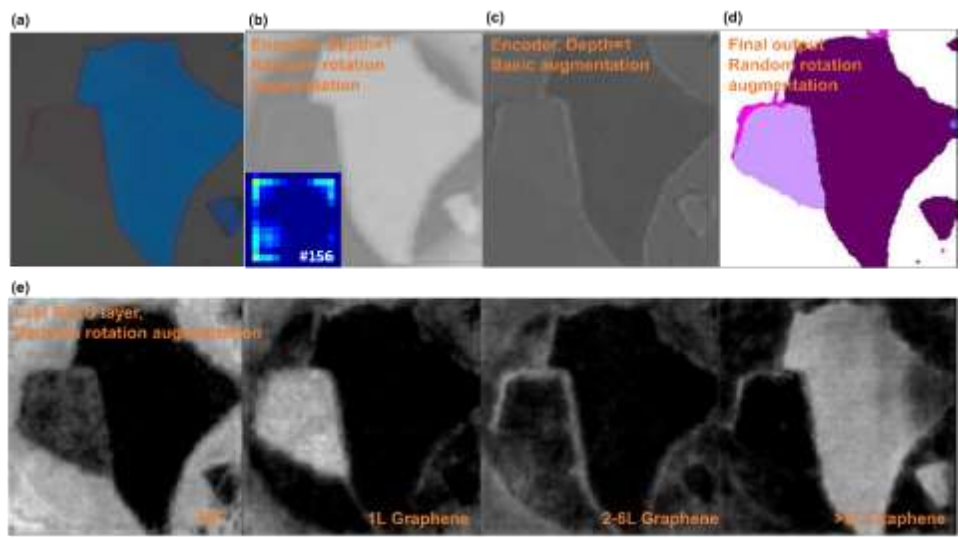

**Figure S52.** Circular shape artifacts induced by the random rotation data augmentation. (A) the input OM image. (b) Feature map in Depth=1 encoder when using the random rotation augmentation method, a clear circle region can be noticed in this map, which is resulted from the padding procedure after image rotation. The inset is the feature map of channel #156 in the Depth=5 encoder layer. (c) Feature map in Depth = 1 encoder using the basic augmentation method. No circle region can be observed in this map. (d) The final predicted label map using random rotation augmentation method, the influence of padding is eliminated as no obvious prediction error happens outside the round region. (e) the final ReLU layer output of the network using

random rotation augmentation method, these four images represent background, Graphene monolayer, Graphene fewlayer and Graphene multilayer.

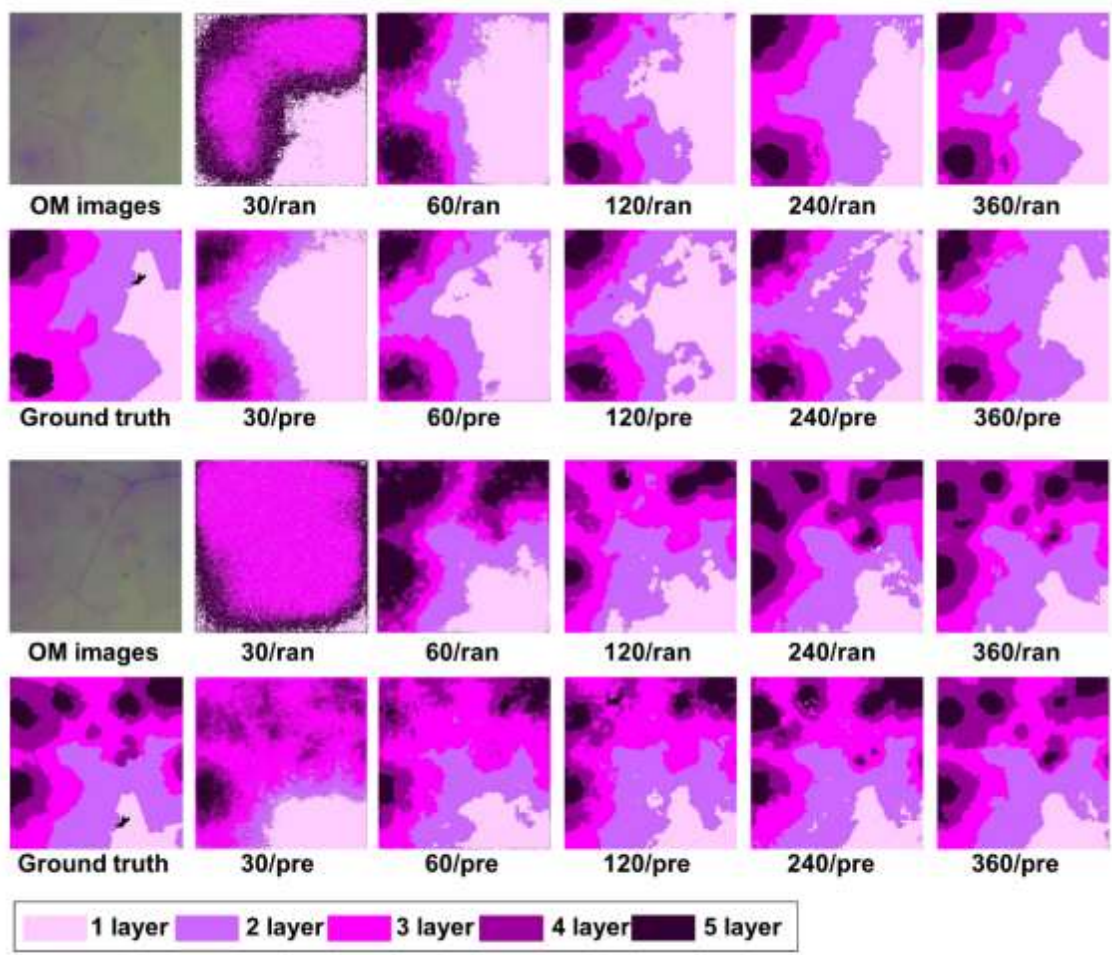

**Figure S53.** Transfer learning results for CVD graphene. The number of training images are varied from 30 to 360 for both the pretraining method (pre) and the random initialization (ran).

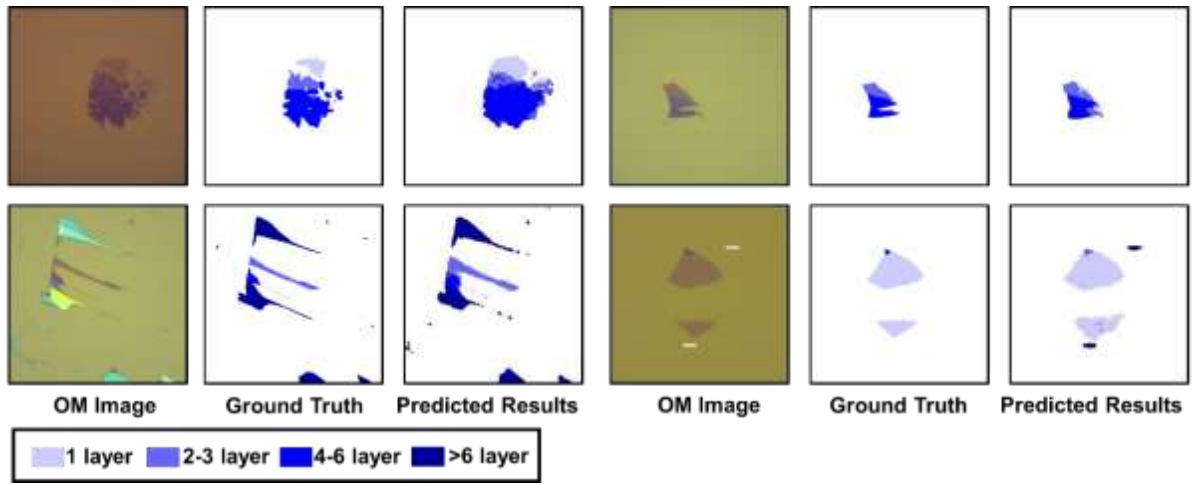

**Figure S54.** Transfer learning results for exfoliated Td-WTe$_2$.

## Reference

1. Long J, Shelhamer E, Darrell T. Fully convolutional networks for semantic segmentation.  2015 IEEE Conference on Computer Vision and Pattern Recognition (CVPR); 2015 7-12 June 2015; 2015. p. 3431-3440.

2. Chen L, Papandreou G, Kokkinos I, Murphy K, Yuille AL. DeepLab: Semantic Image Segmentation with Deep Convolutional Nets, Atrous Convolution, and Fully Connected CRFs. *IEEE Transactions on Pattern Analysis and Machine Intelligence* 2018, **40**(4)**:** 834-848.

3. Ronneberger O, Fischer P, Brox T. U-Net: Convolutional Networks for Biomedical Image Segmentation. In: Navab N, Hornegger J, Wells WM, Frangi AF, editors. Medical Image Computing and Computer-Assisted Intervention – MICCAI 2015; 2015 2015//; Cham: Springer International Publishing; 2015. p. 234-241.

4. Badrinarayanan V, Kendall A, Cipolla R. SegNet: A Deep Convolutional Encoder-Decoder Architecture for Image Segmentation. *IEEE Transactions on Pattern Analysis and Machine Intelligence* 2017, **39**(12)**:** 2481-2495.

5. Geim AK, Novoselov KS. The rise of graphene. *Nature Materials* 2007, **6:** 183.

6.  Ferrari AC, Bonaccorso F, Fal'ko V, Novoselov KS, Roche S, Bøggild P, *et al.* Science and technology roadmap for graphene, related two-dimensional crystals, and hybrid systems. *Nanoscale* 2015, **7**(11)**:** 4598-4810.

7.  Nicolosi V, Chhowalla M, Kanatzidis MG, Strano MS, Coleman JN. Liquid Exfoliation of Layered Materials. *Science* 2013, **340**(6139)**:** 1226419.

8.  Wang QH, Kalantar-Zadeh K, Kis A, Coleman JN, Strano MS. Electronics and optoelectronics of two-dimensional transition metal dichalcogenides. *Nature Nanotechnology* 2012, **7:** 699.

9.  Tan C, Cao X, Wu X-J, He Q, Yang J, Zhang X*, et al.* Recent Advances in Ultrathin Two-Dimensional Nanomaterials. *Chemical Reviews* 2017, **117**(9)**:** 6225-6331.

10. Novoselov KS, Geim AK, Morozov SV, Jiang D, Zhang Y, Dubonos SV*, et al.* Electric Field Effect in Atomically Thin Carbon Films. *Science* 2004, **306**(5696)**:** 666-669.

11. Yi M, Shen Z. A review on mechanical exfoliation for the scalable production of graphene. *Journal of Materials Chemistry A* 2015, **3**(22)**:** 11700-11715.

12. Masubuchi S, Morimoto M, Morikawa S, Onodera M, Asakawa Y, Watanabe K*, et al.* Autonomous robotic searching and assembly of two-dimensional crystals to build van der Waals superlattices. *Nature Communications* 2018, **9**(1)**:** 1413.

13. Li H, Wu J, Huang X, Lu G, Yang J, Lu X*, et al.* Rapid and Reliable Thickness Identification of Two-Dimensional Nanosheets Using Optical Microscopy. *ACS Nano* 2013, **7**(11)**:** 10344-10353.

14. Lin X, Si Z, Fu W, Yang J, Guo S, Cao Y*, et al.* Intelligent identification of two-dimensional nanostructures by machine-learning optical microscopy. *Nano Research* 2018, **11**(12)**:** 6316-6324.

15. Masubuchi S, Machida T. Classifying optical microscope images of exfoliated graphene flakes by data-driven machine learning. *npj 2D Materials and Applications* 2019, **3**(1)**:** 4.

16. Ni ZH, Wang HM, Kasim J, Fan HM, Yu T, Wu YH*, et al.* Graphene Thickness Determination Using Reflection and Contrast Spectroscopy. *Nano Letters* 2007, **7**(9)**:** 2758-2763.

17. Nolen CM, Denina G, Teweldebrhan D, Bhanu B, Balandin AA. High-Throughput Large-Area Automated Identification and Quality Control of Graphene and Few-Layer Graphene Films. *ACS Nano* 2011, **5**(2)**:** 914-922.

18. Blake P, Hill EW, Castro Neto AH, Novoselov KS, Jiang D, Yang R, *et al.* Making graphene visible. *Applied Physics Letters* 2007, **91**(6)**:** 063124.

19. [cited]Available from: https://www.2dsemiconductors.com/; http://www.hqgraphene.com/

20. Sutskever I, Martens J, Dahl G, Hinton G. On the importance of initialization and momentum in deep learning.  International conference on machine learning; 2013; 2013. p. 1139-1147.

21. Simonyan K, Zisserman A. Very deep convolutional networks for large-scale image recognition. *arXiv preprint arXiv:14091556* 2014.

22. Szegedy C, Wei L, Yangqing J, Sermanet P, Reed S, Anguelov D, *et al.* Going deeper with convolutions.  2015 IEEE Conference on Computer Vision and Pattern Recognition (CVPR); 2015 7-12 June 2015; 2015. p. 1-9.

23. Boykov YY, Jolly M. Interactive graph cuts for optimal boundary & region segmentation of objects in N-D images.  Proceedings Eighth IEEE International Conference on Computer Vision. ICCV 2001; 2001 7-14 July 2001; 2001. p. 105-112 vol.101.

24. Wilson JA, Yoffe AD. The transition metal dichalcogenides discussion and interpretation of the observed optical, electrical and structural properties. *Advances in Physics* 1969, **18**(73)**:** 193-335.

25. Beal AR, Hughes HP, Liang WY. The reflectivity spectra of some group VA transition metal dichalcogenides. *Journal of Physics C: Solid State Physics* 1975, **8**(24)**:** 4236-4234.

26. McGuire MA, Clark G, Kc S, Chance WM, Jellison GE, Cooper VR, *et al.* Magnetic behavior and spin-lattice coupling in cleavable van der Waals layered CrCl3 crystals. *Physical Review Materials* 2017, **1**(1)**:** 014001.

27. Lado JL, Fernández-Rossier J. On the origin of magnetic anisotropy in two dimensional CrI3. *2D Materials* 2017, **4**(3)**:** 035002.

28. Sinn S, Kim CH, Kim BH, Lee KD, Won CJ, Oh JS, *et al.* Electronic Structure of the Kitaev Material α-RuCl3 Probed by Photoemission and Inverse Photoemission Spectroscopies. *Scientific Reports* 2016, **6:** 39544.

29. Xiong H, Sobota JA, Yang SL, Soifer H, Gauthier A, Lu MH, *et al.* Three-dimensional nature of the band structure of ZrTe5 measured by high-momentum-resolution photoemission spectroscopy. *Physical Review B* 2017, **95**(19)**:** 195119.

30. Guizzetti G, Nosenzo L, Pollini I, Reguzzoni E, Samoggia G, Spinolo G. Reflectance and thermoreflectance studies of CrCl3, CrBr3, NiCl2, and NiBr2 crystals. *Physical Review B* 1976, **14**(10)**:** 4622-4629.

31. Zhang K, Deng K, Li J, Zhang H, Yao W, Denlinger J*, et al.* Widely tunable band gap in a multivalley semiconductor SnSe by potassium doping. *Physical Review Materials* 2018, **2**(5)**:** 054603.

32. Burton LA, Whittles TJ, Hesp D, Linhart WM, Skelton JM, Hou B*, et al.* Electronic and optical properties of single crystal SnS2: an earth-abundant disulfide photocatalyst. *Journal of Materials Chemistry A* 2016, **4**(4)**:** 1312-1318.

33. Evans BL, Hazelwood RA. Optical and electrical properties of SnSe2. *Journal of Physics D: Applied Physics* 1969, **2**(11)**:** 1507-1516.

34. Molina-Mendoza AJ, Barawi M, Biele R, Flores E, Ares JR, Sánchez C*, et al.* Electronic Bandgap and Exciton Binding Energy of Layered Semiconductor TiS3. *Advanced Electronic Materials* 2015, **1**(9)**:** 1500126.

35. Kurita S, Staehli JL, Guzzi M, Lévy F. Optical properties of ZrS3 and ZrSe3. *Physica B+C* 1981, **105**(1)**:** 169-173.

36. Du K-z, Wang X-z, Liu Y, Hu P, Utama MIB, Gan CK*, et al.* Weak Van der Waals Stacking, Wide-Range Band Gap, and Raman Study on Ultrathin Layers of Metal Phosphorus Trichalcogenides. *ACS Nano* 2016, **10**(2)**:** 1738-1743.

37. Grasso V, Neri F, Perillo P, Silipigni L, Piacentini M. Optical-absorption spectra of crystal-field transitions in MnPS3 at low temperatures. *Physical Review B* 1991, **44**(20)**:** 11060-11066.

38. Autès G, Isaeva A, Moreschini L, Johannsen JC, Pisoni A, Mori R*, et al.* A novel quasi-one-dimensional topological insulator in bismuth iodide β-Bi4I4. *Nature Materials* 2015, **15:** 154.

39. Choudhary K, Cheon G, Reed E, Tavazza F. Elastic properties of bulk and low-dimensional materials using van der Waals density functional. *Physical Review B* 2018, **98**(1)**:** 014107.

40. Zhou J, Shen L, Costa MD, Persson KA, Ong SP, Huck P*, et al.* 2DMatPedia: An open computational database of two-dimensional materials from top-down and bottom-up approaches. *arXiv preprint arXiv:190109487* 2019.